# Non-volatile electrically programmable integrated photonics with a 5-bit operation

Rui Chen[1,*], Zhuoran Fang[1], Christopher Perez[3], Forrest Miller[1], Khushboo Kumari[1], Abhi Saxena[1], Jiajiu Zheng[1], Sarah J. Geiger[4], Kenneth E. Goodson[3], Arka Majumdar[1,2,*]

[1]Department of Electrical and Computer Engineering, University of Washington, Seattle, WA 98195, USA

[2]Department of Physics, University of Washington, Seattle, WA 98195, USA

[3]Department of Mechanical Engineering, Stanford University, Stanford, CA 94305, United States

[4]The Charles Stark Draper Laboratory, Cambridge, MA 02139, USA

*Email: charey@uw.edu and arka@uw.edu

# Abstract

Scalable programmable photonic integrated circuits (PICs) can potentially transform the current state of classical and quantum optical information processing. However, traditional means of programming, including thermo-optic, free carrier dispersion, and Pockels effect result in either large device footprints or high static energy consumptions, significantly limiting their scalability. While chalcogenide-based non-volatile phase-change materials (PCMs) could mitigate these problems thanks to their strong index modulation and zero static power consumption, they often suffer from large absorptive loss, low cyclability, and lack of multilevel operation. Here, we report a wide-bandgap PCM antimony sulfide ($Sb_2S_3$)-clad silicon photonic platform simultaneously achieving low loss, high cyclability, and 5-bit operation. We switch $Sb_2S_3$ via an on-chip silicon PIN diode heater and demonstrate components with low insertion loss (<1.0 dB), high extinction ratio (>10 dB), and high endurance (>1,600 switching events). Remarkably, we find that $Sb_2S_3$ can be programmed into fine intermediate states by applying identical and thermally isolated pulses, providing a unique approach to controllable multilevel operation. Through dynamic pulse control, we achieve on-demand and accurate 5-bit (32 levels) operations, rendering 0.50 ± 0.16 dB contrast per step. Using this multilevel behavior, we further trim random phase error in a balanced Mach-Zehnder interferometer. Our work opens an attractive pathway toward non-volatile large-scale programmable PICs with low-loss and on-demand multi-bit operations.

# Main

Programmable photonic integrated circuits (PICs), usually composed of arrays of tunable beam splitters and phase shifters, can change their functionalities on demand. This flexibility has recently extended their traditional applications from optical interconnects to optical computing[1], optical programmable gate arrays[2], and quantum information processing[3]. Further scaling of the programmable PICs requires the constituent devices to have a smaller footprint and lower power consumption. However, current programmable PICs are primarily based on weak volatile tuning mechanisms, such as thermo-optic effects[4,5], free-carrier effects[6], and Pockels effects[7], which provide a small change in refractive index ($\Delta n < 0.01$). Therefore, the resulting devices usually suffer from large footprints (> 100 μm), and require a constant power supply (~10 mW[1]). Moreover, thermo-optic effects incur severe thermal crosstalk, necessitating additional heaters and control circuits for crosstalk compensation. These requirements severely limit the current PIC's integration density and energy efficiency.

Chalcogenide-based non-volatile phase-change materials (PCMs) can potentially alleviate these issues[8–10]. PCMs have two stable, reversibly switchable micro-structural phases (amorphous and crystalline, termed here as a- and c-phase), with drastically different optical refractive indices ($\Delta n \sim 1$). Thanks to the non-volatile phase transition, no static power is needed to hold the state upon switching PCMs' phase. The substantial refractive index change $\Delta n$ and nonvolatility enable compact reconfigurable devices (~10 μm[10,11]) with zero static energy consumption[12–17]. Moreover, PCMs are compatible with large-scale integration since they can be easily deposited by sputtering[12,15,17–19] or thermal evaporation[11] onto almost any PIC materials, including silicon and silicon nitride. Despite these advantages, archetypal PCMs such as $Ge_2Sb_2Te_5$ (GST) and GeTe exhibit strong absorption in both phases at relevant optical communication wavelengths. This hinders their utility in phase shifters – an essential building block for programmable PICs. Emerging wide-bandgap PCMs, such as GeSbSeTe (GSST)[20], antimony selenide ($Sb_2Se_3$)[21], and antimony sulfide ($Sb_2S_3$)[22] can circumvent this loss and have recently generated strong interest in the community[11,16,23–25]. In particular, $Sb_2S_3$ shows the widest bandgap among all these PCMs, allowing transparency down to ~ 600 nm in the amorphous phase[22]. Moreover, the lack of selenium in $Sb_2S_3$ makes it less toxic[26] and less prone to cause chamber contamination during sputtering or evaporation processes. Thus, $Sb_2S_3$ is much more amenable to be adapted in a commercial foundry. Despite these promises, high endurance electrical control of $Sb_2S_3$ remained unsolved.

Here, we demonstrate electrically controlled $Sb_2S_3$-clad silicon photonic devices with low insertion loss (<1dB), high extinction ratio (>10dB), and high endurance (>1,600 switching events). The phase transition is actuated by silicon PIN (p-doped-intrinsic-n-doped) diode heaters. We established the versatility of this hybrid platform with three different integrated photonic devices, including microring resonators, Mach-Zehnder interferometers (MZIs), and asymmetric directional couplers. We also observed multilevel (32 levels) operation, highest among all reported electrically controlled PCM-silicon platforms, with a resolution of 0.50 ± 0.16 dB per step. This multilevel operation is achieved by sending multiple thermally separated (~1 second) near-identical electrical pulses for both amorphization and crystallization. We leveraged this multilevel behavior to trim a balanced MZI to correct the random phase error caused by fabrication imperfections. The multilevel operation is crucial to avoid under or over-correction during trimming. Our work opens a new avenue for non-volatile electrically programmable PICs with low loss and on-demand multilevel operation.

## Results

We characterized the sputtered $Sb_2S_3$ thin films to obtain the refractive indices and verify the crystallization capability (annealed under 325° for 10 mins) of the as-deposited a-$Sb_2S_3$ (Supplementary Section S1). The silicon photonic devices (Fig. 1-Fig. 3) were designed to operate at the telecommunication O-band (1260 - 1360 nm) and fabricated on a standard silicon-on-insulator (SOI) wafer with 220 nm silicon and 2 μm buried oxide. The 500-nm-wide waveguides are fabricated by partially etching 120-nm silicon. We then deposited 450-nm-wide $Sb_2S_3$ onto the SOI chip via sputtering. The slightly smaller width than the waveguide is to compensate for electron beam lithography (E-beam) overlay tolerance. Our simulation results show a change in effective index $\Delta n_{eff} \approx 0.018$ between a- and c-$Sb_2S_3$ (Supplementary Section S2). The $Sb_2S_3$ films are electrical controlled via on-chip silicon PIN micro-heaters[15,17]. The $p^{++}$ and $n^{++}$ doping regions were designed 200 nm away from the waveguide to avoid free-carrier absorption loss[17], indicated in the scanning electron microscope (SEM) images with false colors in Fig. 1c, Fig. 2c, and Fig. 3c. The $Sb_2S_3$ stripes are encapsulated with 40 nm of $Al_2O_3$ grown by atomic layer deposition (ALD) under 150°C. This conformal encapsulation is critical to prevent $Sb_2S_3$ from oxidation and thermal reflowing, and thus is essential to attain high endurance. To show that our $Sb_2S_3$-clad silicon photonic platform is versatile and compatible with most PIC components, we demonstrate three widely used PIC components: (1) a microring resonator to show low-loss tuning of cavities, (2) a balanced MZI to demonstrate a full π phase shift and a broadband operation, and (3) an asymmetric directional coupler to create a compact programmable unit (see simulation results in Supplementary Section S2).

**Nonvolatile microring switch integrated with $Sb_2S_3$ phase shifter**

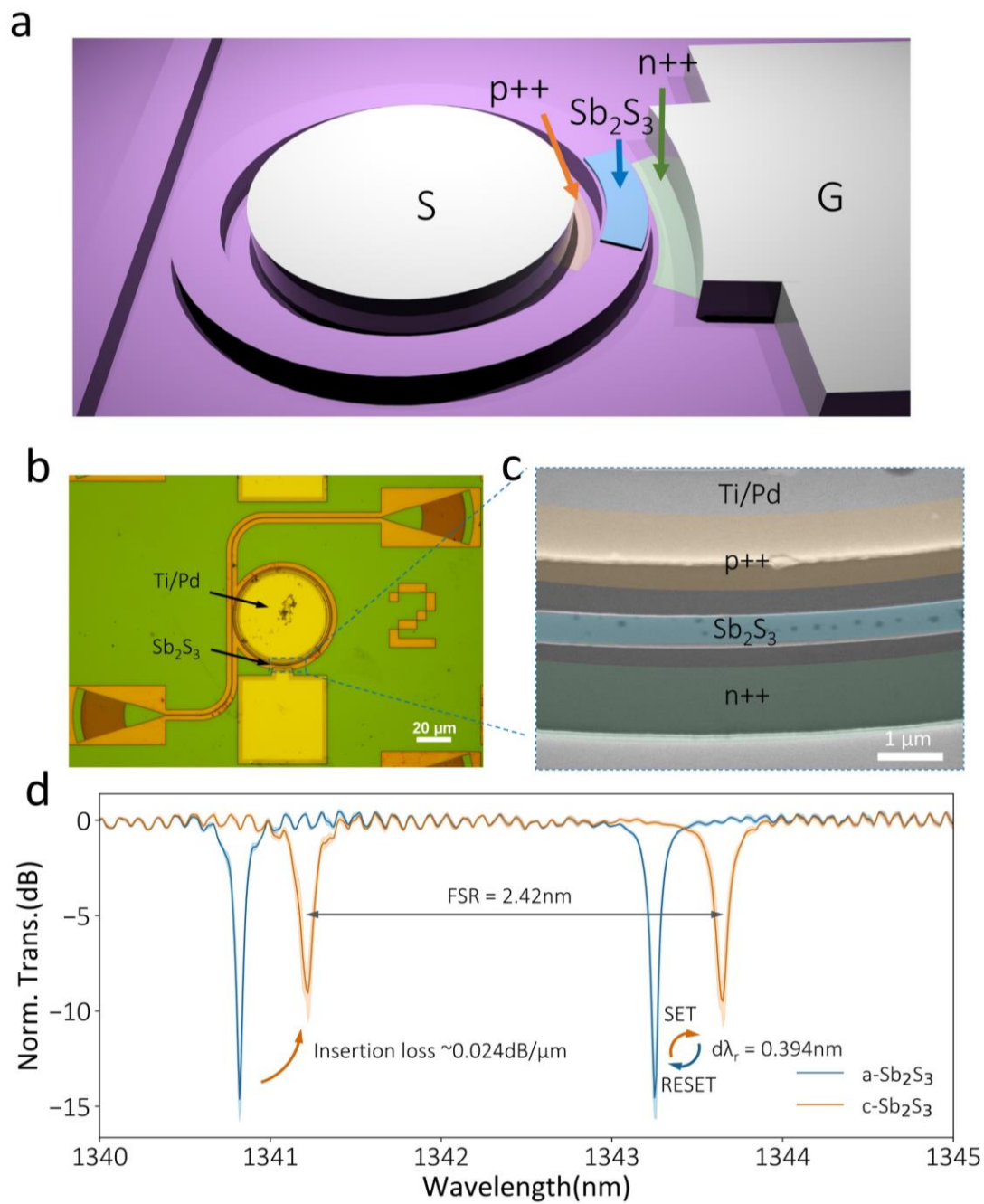

**Fig. 1: A high-Q ring resonator loaded with 10 µm of 20-nm-thick $Sb_2S_3$.** (a) Schematic of the device (the encapsulation ALD $Al_2O_3$ layer is not shown), (b) optical, and (c) Scanning-Electron Microscope (SEM) images of the micro-ring resonator. The $Sb_2S_3$ thin film, $n^{++}$, and $p^{++}$ doped silicon regions are represented by false colors (blue, orange, and green, respectively). (d) Measured micro-ring spectra in two phases (SET: three 1.6 V, 200-ms-long pulses to change $Sb_2S_3$ to the crystalline state; RESET: three 7.5 V, 150-ns-short pulses to change $Sb_2S_3$ to the amorphous state). The spectra are averaged over ten cycles of reversible electrical switching, and the shaded area shows the standard deviation. Norm. Trans. Stands for normalized transmission, normalized to a reference waveguide on the same chip.

We deposited 10-µm-long, 20-nm-thick $Sb_2S_3$ on a micro-ring resonator with 30 µm radius (Figs. 1a-c). The free spectral range (FSR) is ~2.42 nm (Fig. 1d), and the bus-ring gap is 280 nm to achieve a near-critically coupled device. We switched the as-deposited a-$Sb_2S_3$ to c-$Sb_2S_3$ on the microring resonator by applying three 1.6 V, 200-ms-long pulses (or SET pulses) separated by 1 second and then re-amorphized the material via three 7.5 V, 150-ns-short pulses also separated by 1 second (RESET pulses). We note that the 200-ms-long SET pulse is indeed significantly longer than other reported PCMs, such as GST (50 µs[17] or 100 µs[16]) and $Sb_2Se_3$ (5 µs[11] or 100 µs[16]). But we found that the SET pulse duration could be further reduced to around 100 µs after the first few cycles (see Method and Supplementary Section S12). Moreover, the slow crystallization allows a large volume of $Sb_2S_3$ to amorphize (Supplementary Section S13). Three pulses instead of a single pulse were used to ensure complete amorphization and crystallization (see Methods). Fig. 1d shows a resonance shift of ~0.394 nm upon switching the 10 µm $Sb_2S_3$ from the a- (blue) to the c-phase (orange), corresponding to a π-phase shift length $L_\pi$ of ~30.7 µm, significantly shorter than the 1-millimeter $L_\pi$ of ferroelectric non-volatile phase shifter[27]. The SET and RESET processes were repeated for 10 cycles. The shift in resonance is highly repeatable, as suggested by the slight standard deviation (Fig. 1d). The excess loss from a-$Sb_2S_3$ is negligible[19], and in c-$Sb_2S_3$ hybrid waveguides the loss is estimated to be 0.024 dB/µm (0.72 dB/π), which is three times larger than our simulation result (0.26 dB/π) (Supplementary Section S2). We attribute this excess loss to the scattering from the $Sb_2S_3$ thin film due to non-uniform deposition/liftoff and local crystal grains in c-$Sb_2S_3$ (Supplementary Section S3). We verified that the loss due to mode mismatch at the transition between the bare silicon waveguide and the $Sb_2S_3$-loaded waveguide is small (~0.013 dB/facet), consistent with the fact that the thin $Sb_2S_3$ film should not significantly change the mode shape (Supplementary Section S2).

**Nonvolatile Mach-Zehnder switch integrated with $Sb_2S_3$ phase shifter**

Figs. 2a-c show a balanced MZI operating at wavelengths between 1,320 nm and 1,360 nm with both arms covered with 30-μm-long 20-nm-thick $Sb_2S_3$. A multimode interferometer with a 50:50 splitting ratio was designed and fabricated, as shown in Fig. 2d (simulation results in Supplementary Section S2). Initially, the light comes out mainly from the bar port with an extinction ratio of ~13 dB (Fig. 2e). The light coming out from the bar port instead of the cross port in this balanced MZI can be explained by the random phase errors in two arms due to fabrication imperfection, especially that in the S-bend. One can overcome such imperfection by exploiting a wider waveguide to improve fabrication robustness[28]. Alternatively, this random phase error can be corrected using $Sb_2S_3$ for post-fabrication trimming, as we show later in this paper. The $Sb_2S_3$ on one arm was switched by two 1.7 V, 200-ms SET electric pulses to provide a full π phase shift. An 8.1 V, 150-ns short RESET electric pulse switched the device back to the initial state. Figs. 2e and 2f show the transmission spectra normalized to a reference waveguide when the $Sb_2S_3$ film is in the amorphous and crystalline phases, respectively. The c-$Sb_2S_3$ displays a complete spectrum flip, showing a bar state with an extinction ratio of 15 dB. We then recorded the bar port transmission at 1,330 nm for 100 switching events without device degradation (Supplementary Section S4).

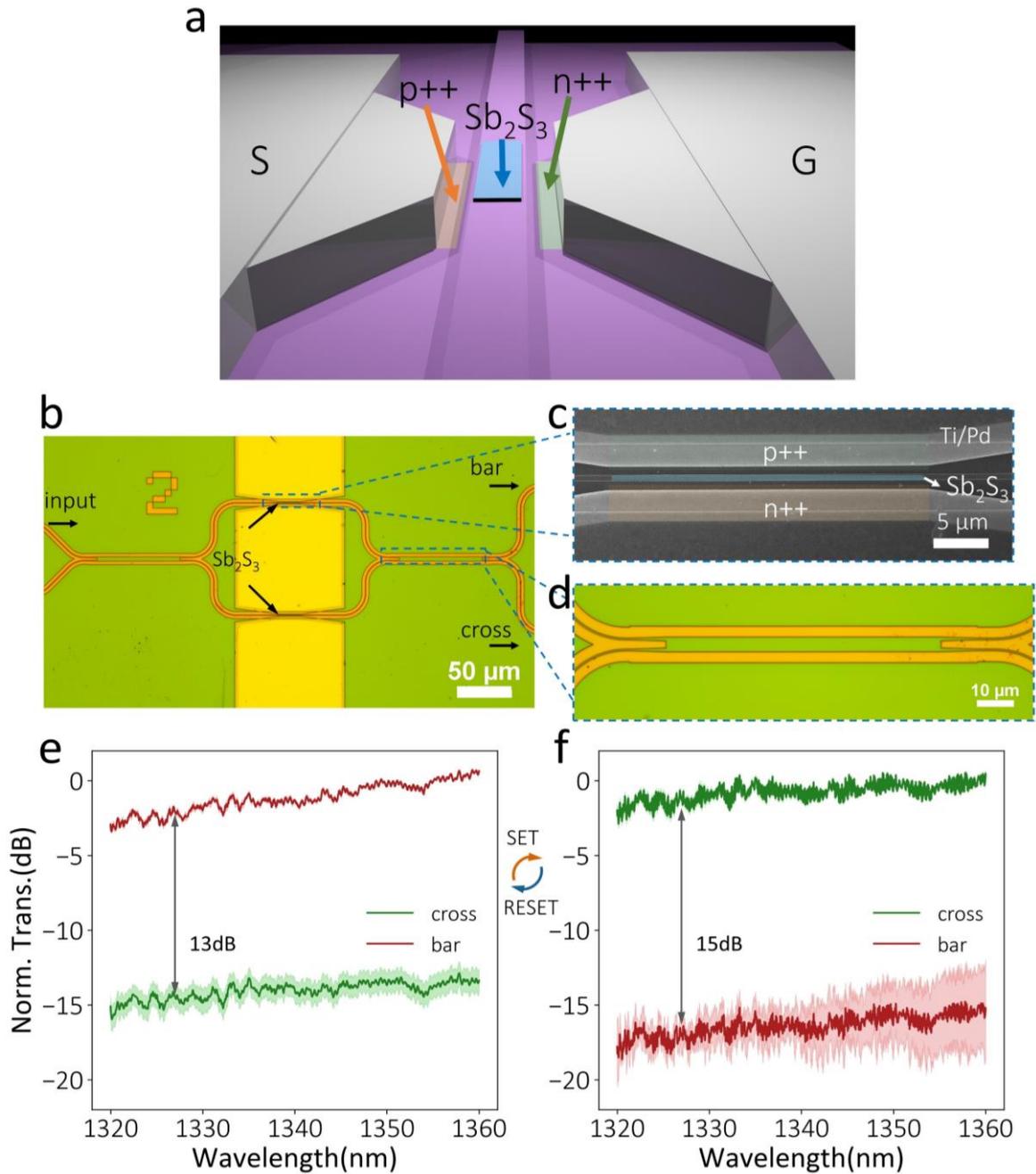

**Fig. 2: A Mach-Zehnder interferometer with both arms covered with 20-nm-thick Sb$_2$S$_3$.** (a) Sb$_2$S$_3$ phase shifter schematic (the encapsulation ALD Al$_2$O$_3$ layer is not shown). (b) Optical and (c) SEM images of the Sb$_2$S$_3$ phase shifter. (d) Optical micrograph of the 50:50 splitting ratio multi-mode interferometer. (e, f) Transmission spectra at both bar and cross ports for (e) a-Sb$_2$S$_3$ (RESET: an 8.1 V, 150 ns pulse) and (f) c-Sb$_2$S$_3$ (SET: two 1.7 V, 200 ms pulses). The green and red lines represent the measured transmission at cross and bar ports. The shaded region indicates the standard deviation of transmission for 5 cycles of measurements.

**Compact asymmetric directional coupler switch**

We also designed and fabricated a compact asymmetric directional coupler (coupling length $L_c \approx 79$ µm) (simulation in Supplementary Section S2), as shown in Figs. 3a-c. The coupler consists of two waveguides with different widths. The narrower 409-nm-wide waveguide (hybrid waveguide) was capped with 20-nm-thick $Sb_2S_3$ and designed to allow phase match with the wider 450-nm-wide waveguide (bare waveguide) for c-$Sb_2S_3$[29]. As such, the input light could completely couple to the cross port in one coupling length. Once $Sb_2S_3$ is switched to the amorphous state, the effective index of the hybrid waveguide changes while the bare waveguide remains the same. The resulting phase mismatch changes the coupling strength and coupling length. Then, co-optimizing the gap and waveguide length permits a complete bar transmission. The unique c-$Sb_2S_3$ phase matching approach, instead of a-$Sb_2S_3$[13,15,29,30], allows a more symmetric performance regardless of the input port (Supplementary Section S2), crucial for a $2 \times 2$ device. If phase mismatch happens in the c-$Sb_2S_3$ state, the slight loss of c-$Sb_2S_3$ on one of the waveguides will result in different bar state insertion loss when the light goes from different input ports. We note that the 79-µm coupling length can be potentially reduced (~ 34 µm) by depositing a thicker (50 nm) $Sb_2S_3$ to provide stronger refractive index modulation.

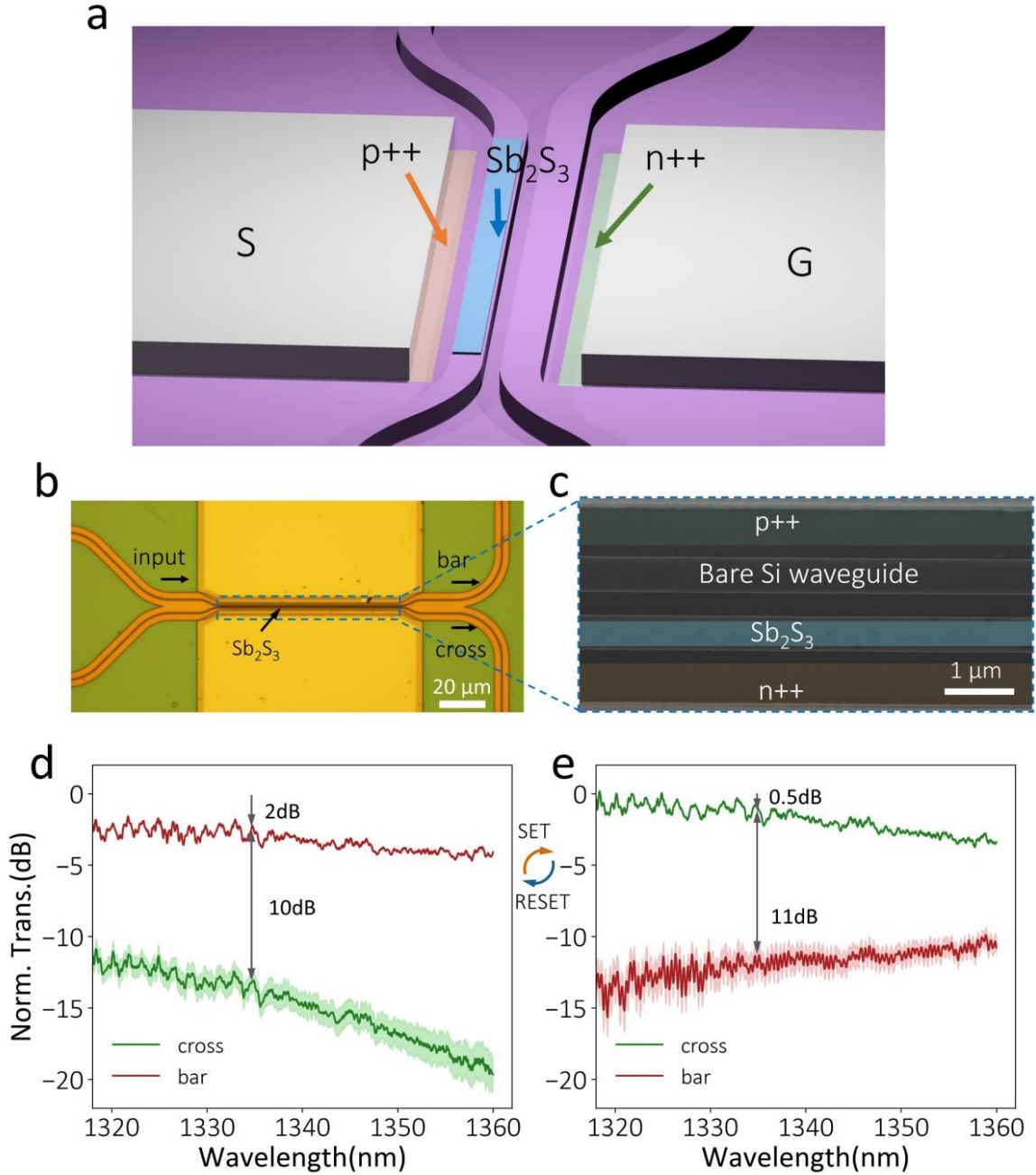

**Fig. 3: An asymmetric directional coupler with $Sb_2S_3$-Si hybrid waveguide.** (a) Schematic (the encapsulation ALD $Al_2O_3$ layer is not shown), (b) optical, and (c) SEM images of the asymmetric directional coupler. (d, e) Transmission spectra at bar and cross ports for (d) a-$Sb_2S_3$ (RESET: three 9.6 V, 500 ns pulses) and (e) c-$Sb_2S_3$ (SET: three 2.7 V, 200 ms pulses). The result was averaged over five measurements, and the shaded region indicates the standard deviation.

Figs. 3d and 3e show the transmission spectra for a- and c-$Sb_2S_3$, switched with three 9.6 V, 500 ns RESET, and 2.7 V, 200 ms SET pulses, respectively. The insertion losses are 2 dB (0.5 dB), and the extinction ratios are around 10 dB (11 dB) for a (c)-$Sb_2S_3$). The unexpected high insertion loss when the $Sb_2S_3$ is in the amorphous state can be attributed to several factors, including gap discrepancy, $Sb_2S_3$ overlay deviation, or the cross-port grating coupler fabrication imperfection. To estimate the actual loss of the device, we apply the c-$Sb_2S_3$ loss extracted from the ring resonator to the simulation and calculate this device's insertion loss to be ~0.1 dB (0.9 dB) for a(c)-$Sb_2S_3$ (Supplementary Section S2).

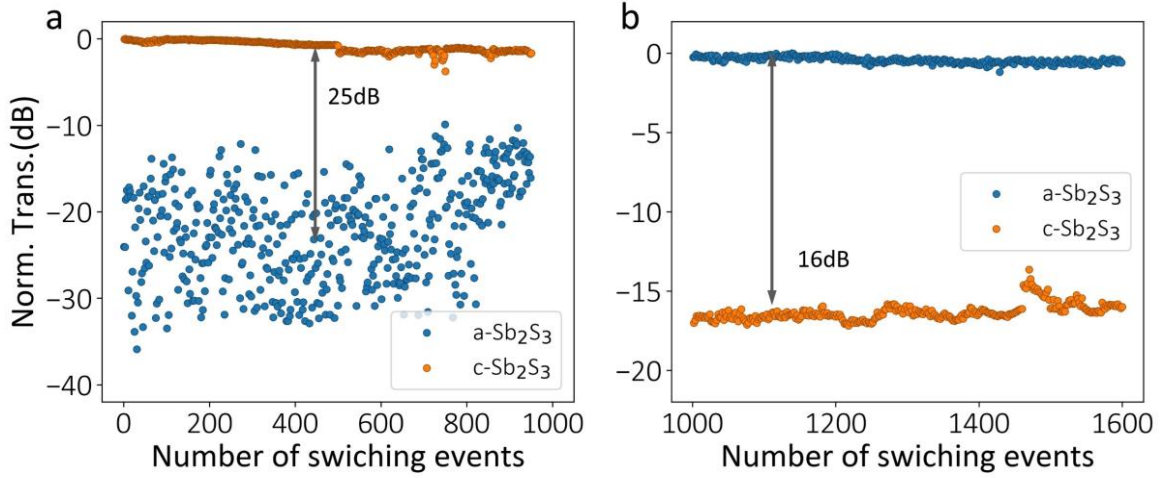

**Fig. 4: Cyclability test for a-$Sb_2S_3$-based asymmetric directional coupler.** Measured transmission at the (a) cross and (b) bar ports. The blue and orange scatterers represent the normalized transmission when $Sb_2S_3$ is in the amorphous and crystalline phases. The phase change condition is the same as in Fig.3. The device was switched for over 1,600 events with no significant insertion loss and performance degradation.

Fig. 4 shows 1,600 switching events for the asymmetric directional coupler. Limited by our measurement setup, we separately measured the cross (Fig. 4a) and bar ports (Fig. 4b). The higher insertion loss (~1 dB) at around event 500 was due to optical fiber misalignment. We note that almost no performance degradation occurred at the end (Supplementary Section S5); hence, 1,600 switching events are not the limit of this device. The cross-port transmission shows a relatively large variation for a-$Sb_2S_3$ (Fig. 4a blue scatterers, from ~ -15 dB to ~ -35 dB), which was caused by incomplete amorphization or thermal reflow of the $Sb_2S_3$ film. Since the plot is on a logarithmic scale, such a large variation in a-$Sb_2S_3$ (due to higher transmission) is not visible in Fig. 4b.

## Multilevel 5-bit operation with dynamic electrical control

Our $Sb_2S_3$-Si integrated structures further show a stepwise multilevel operation up to 32 levels with dynamic pulse control. In Fig. 5a, we show the multilevel transmission at both cross and bar ports of an asymmetric directional coupler while sending in RESET 10 V, 550 ns pulse every other second to amorphized the $Sb_2S_3$. We started the experiment with a "coarse" tuning, where unoptimized, identical pulses were sent to partially amorphize the c-$Sb_2S_3$ device to demonstrate multiple levels. The asymmetric directional coupler was originally in the "cross state" (red region). After one partial amorphization pulse, it was reconfigured into an intermediate state (orange region), where light comes out from both cross and bar ports. After six pulses, a complete "bar state" (green region) was achieved. We repeated this experiment five times for each port and plotted the average transmission levels and the standard deviation. The variation is attributed to the stochastic phase change process using electrical controls[31]. We also experimentally tested the partial crystallization of the device and multilevel operation (Supplementary Section S5), which is based on the growth-dominant nature of $Sb_2S_3$ crystallization process[29]. In the following experiments, we mainly focused on partial amorphization because of lower energy consumption and more operation levels with finer resolution.

Such stepwise multilevel operation by applying identical pulses is distinctly different from previously reported multilevel operations in GST[15,17] and $Sb_2Se_3$[11,16], where different voltage amplitudes or pulse duration were used to access multiple levels during amorphization. The pulse-number-dependent behavior is quite counterintuitive: one expects that after the first amorphization pulse, the thin $Sb_2S_3$ film would have reached its new equilibrium phase. Moreover, since the thermal processes relax within 10 μs (Supplement Section S6), each pulse is independent due to the relatively long one-second interval. As a result, the subsequent identical and separated pulses should not further change the material phase. To understand the origin of the multi-level operation, we closely inspected four partially amorphized $Sb_2S_3$ devices under the microscope. We observed a few separate patches (Supplement Section S7) and a region that grew with more voltage pulses. As reported in some literature, one possibility could have been that a- and c-$Sb_2S_3$ have significantly different thermal conductivities and specific heat capacities. But we measured the thermal conductivities to be similar (a-$Sb_2S_3$: 0.2 W/m/K; c-$Sb_2S_3$: 0.4 W/m/K, see Methods), and hence, we ruled out this as a possible explanation. We hypothesize that this unique behavior comes from $Sb_2S_3$'s multiple crystalline phases. $Sb_2S_3$ has at least two distinct crystalline phases[32], which may differ in the amorphization conditions. The partial amorphization pulse can cause amorphization in the hottest region, but at the lower temperature region, it may cause phase transition to the other crystalline phase. These regions get amorphized in subsequent pulses, resulting in multilevel operation.

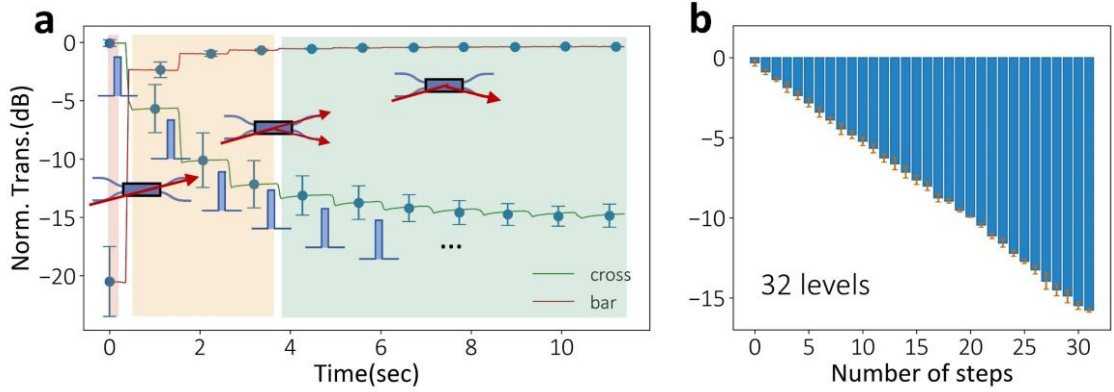

**Fig. 5: A quasi-continuously tunable directional coupler based on multilevel $Sb_2S_3$.** (a) (Coarse tuning applying identical electrical pulses) Time trace measurement of the directional coupler when sending in an amorphization pulse (10V, 550ns) each second. Green (red) curves represent cross (bar) transmission. The result is averaged over five experiments for each port, and the error bar shows the standard deviation. Depending on the pulses, a "bar", "intermediate", and "cross" state can be achieved as indicated by the red, orange, and green regions. (b) (Fine tuning using dynamically controlled near-identical electrical pulses) Normalized transmission at 1,340 nm at the bar port shows 5-bit operation (32 distinct levels), achieved by dynamically controlling the number, amplitude, and duration of pulses sent in (near-identical, 9.65 V ~ 9.85 V, 550 ns). A precise transmission level of 0.50 ± 0.16 dB per step and 32 levels were simultaneously achieved. The only slight difference between the target and achieved transmission demonstrates an on-demand operation. The error bars represent the standard deviation over five experiments.

An even finer multilevel operation was realized by monitoring the transmission level and dynamically changing the pulse conditions slightly. Here, we demonstrate on-demand 5-bit operation in a quasi-continuously tunable directional coupler, as shown in Fig. 5b. We dynamically controlled the partial amorphization pulses to have slightly lower, near-identical voltages (ranging from 9.65 V to 9.85 V) and obtained up to 32 levels. Fig. 5b demonstrates 5-bit operation (32 distinct levels) with a target resolution of 0.5 dB per level step at 1,340 nm (see the detailed pulse conditions in Supplementary Section S8). We emphasize that dynamic control is necessary to mitigate the stochastic nature of electrically controlled PCMs[31], hence essential for a reliable many-level operation. In Fig. 5b, a linear fit shows a slope of -0.50 dB per step and a standard deviation of 0.16 dB among five experiments, indicating a repeatable operation. While the 5-bit operation of GST was shown using laser pulses[33], our demonstration is thus far the highest number of operating levels reported using electrical control in PCMs-based photonics. Moreover, our multilevel operation does not require sophisticated heater geometry engineering, such as the segmented doped silicon heater design[11], and solely relies on the unique phase-change dynamics of $Sb_2S_3$.

## Random phase error correction in balanced MZIs exploiting multilevel operation

Finally, exploiting the multilevel operation, we corrected random phase errors in a balanced MZI. A perfectly balanced MZI should initially be in an all-cross state. However, random phase errors due to fabrication imperfections can easily build up to a phase error of π, making the initial state unpredictable. Therefore, balanced MZIs usually require extra calibration[28]. For example, Fig. 6a shows the transmission spectra of a phase error corrupted balanced MZI, indicating a high bar transmission. Both arms of the MZI are loaded with 40-µm-long $Sb_2S_3$ film to guarantee a phase tuning range of more than ±π. The corrected MZI spectra by multilevel tuning of $Sb_2S_3$ are demonstrated in Fig. 6b, showing a pure "cross state" with a high extinction ratio of 24 dB. The trimming process is shown in Fig. 6c. We sent in a partial amorphization pulse (8.8 V, 150 ns) every other second, which gradually increased the portion of amorphous $Sb_2S_3$, resulting in quasi-continuous changes of the bar (blue) and cross (orange) transmission (at 1,340 nm). The correction finishes once a bar transmission minimum is reached, indicated by the red arrows in Fig. 6c, and the spectra reported in Fig. 6b were then measured. Further pulses increase bar transmission because of the over-compensated phase. We performed the same experiment three times, indicated by different colored regions in Fig. 6c. Complete phase error correction was observed in all three instances, exhibiting excellent repeatability of our trimming process. Note that a binary tuning cannot accomplish this task due to the random initial phase error. Even multilevel operations with limited discrete-level resolution can cause over- or under-corrected phase error, ultimately determining the trimming resolution. We highlight that this method requires zero static energy supply once the phase error is corrected, as the phase transition in $Sb_2S_3$ is non-volatile (> 77 days, Supplementary Section S9). Thanks to the relatively fine operation levels, slight over-tuning does not significantly affect the performance. The trimming resolution can be further improved using dynamically changed, near-identical electrical pulses, as shown in the previous Section. Moreover, if the phase error is over compensated, the device could be tuned back with partial crystallization pulses (Supplementary Section S5). In the future, our trimming process can potentially be fully automated by real-time adjusting the pulse numbers according to the measured transmission.

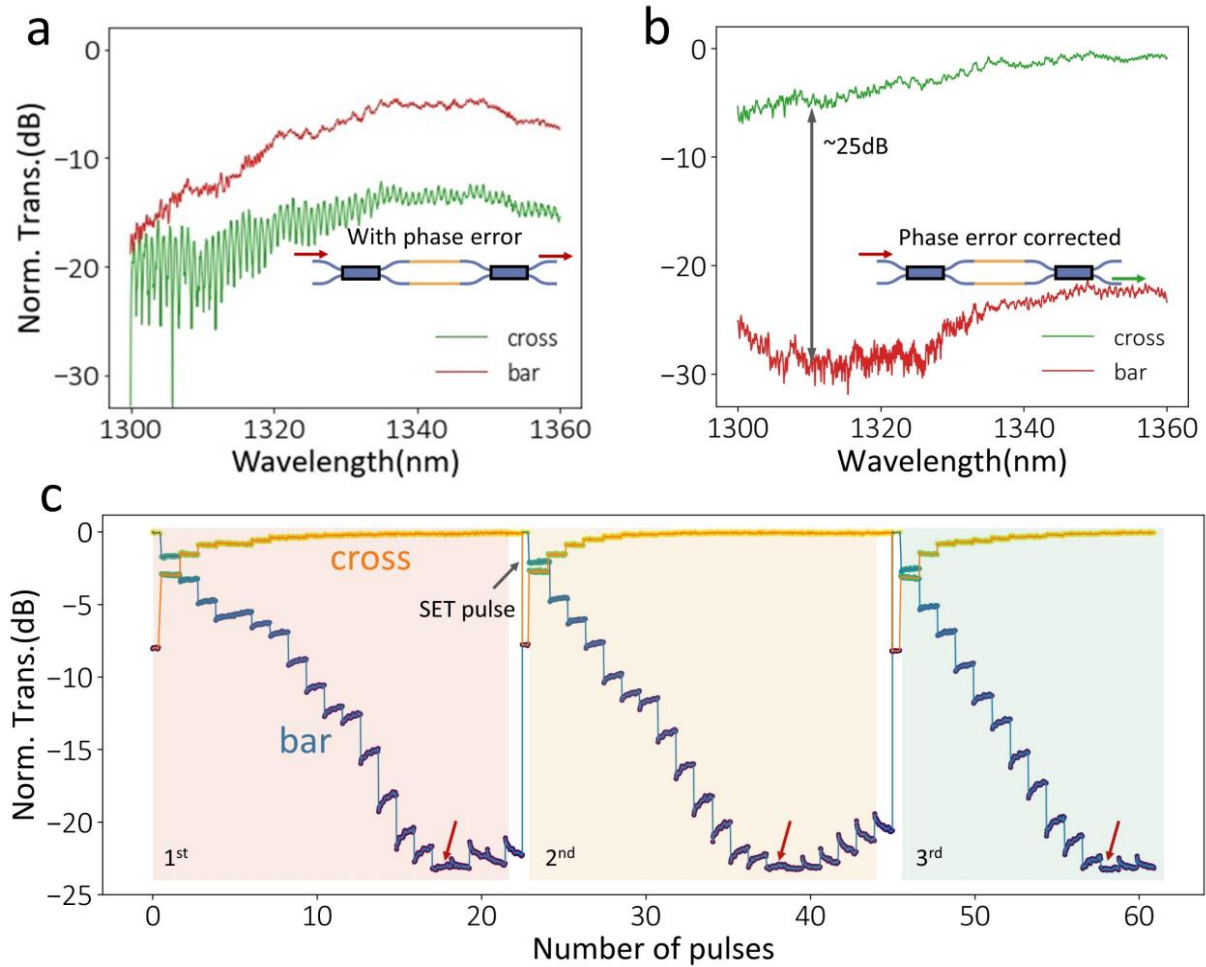

**Fig. 6: Random phase error correction in a balanced MZI based on the multilevel operation.** (a) Transmission spectrum of a balanced MZI with phase error due to fabrication imperfections. The red (green) line represents bar (cross) transmission. Light transmits mostly from the bar port instead of the ideal cross port. (b) Transmission spectrum after the correction with 8.8 V, 150 ns pulses. The device is in a "cross state" with a high extinction ratio of ~24 dB, indicating a phase-error-free device. (c) Time trace measurement for three experiments shows the phase error correction step. The RESET pulses are sent at a one-second interval. As more pulses were sent in, the bar transmission continuously decreased until it reached a minimum of ~-24 dB (red arrow), and then it went up. In all the experiments, the optimal error correction was obtained. The slight volatile increase (only visible when transmission is smaller than -12 dB) after each amorphization pulse could be attributed to the relatively long thermal relaxation time, but the transmission stabilizes before the next pulse. The blue (orange) represents bar- (cross-) transmission.

# Discussion

We demonstrated a multi-bit integrated electro-photonic platform using wide bandgap PCM $Sb_2S_3$ and doped silicon PIN heaters. Operating in the telecommunication O-band, the $Sb_2S_3$-Si hybrid ring resonators, MZIs, and directional couplers exhibit a low insertion loss (< 1.0 dB) and a high extinction ratio (> 10 dB). We report record-high electrical cyclability of $Sb_2S_3$ (> 1,600 switching events). Notably, the $Sb_2S_3$-based photonic devices could be tuned into distinct intermediate levels in a stepwise fashion by consecutively sending identical, thermally isolated partial amorphization pulses. Such pulse-number-dependent multilevel tuning allows high-resolution operation levels compared to different voltages or pulse widths. We demonstrate precise on-demand 5-bit operation with 0.50 ± 0.16 dB per level step in a tunable beam splitter. We further exploited this behavior to demonstrate random phase error correction in a balanced MZI, which will find practical applications. Our work opens a new pathway toward various integrated photonics applications, such as post-fabrication trimming, optical information processing, and optical quantum simulations, where the nonvolatile on-demand multi-bit operation is of paramount interest.

# Methods

*SOI Device Fabrication*

The silicon photonic devices were fabricated on a commercial SOI Wafer with 220-nm-thick Silicon on 2-µm-thick $SiO_2$ (WaferPro). All devices were defined using electron-beam lithography (EBL, JEOL JBX-6300FS) with a positive-tone E-beam resist (ZEP-520A) and partially etched by ~120 nm in a fluorine-based inductively coupled plasma etcher (ICP, Oxford PlasmaLab 100 ICP-18) with mixed $SF_6/C_4F_8$. The etching rate was around 2.8 nm/sec. The doping regions were defined by two additional EBL rounds with 600-nm-thick poly (methyl methacrylate) (PMMA) resist and implanted by boron (phosphorus) ions for $p^{++}$ ($n^{++}$) doping regions with a dosage of $2 \times 10^{15}$ ions per $cm^2$ and ion energy of 14 keV (40 keV). The chips were annealed at 950 °C for 10 min (Expertech CRT200 Anneal Furnace) for dopant activation. Ideal Ohmic contact was formed after removal of the surface native oxide via immersing the chips in 10:1 buffered oxide etchant (BOE) for 10 seconds. The metal contacts were then immediately patterned by a fourth EBL step using PMMA. Metallization was done by electron-beam evaporation (CHA SEC-600) and lift-off of Ti/Pd (5 nm/180 nm). After a fifth EBL defining the $Sb_2S_3$ window, a 40-nm $Sb_2S_3$ thin film was deposited using a GST target (AJA

International) in a magnetron sputtering system (Lesker Lab 18), followed by a lift-off process. We note the actual $Sb_2S_3$ thickness on the waveguide was reduced to around 20nm because of the narrow PMMA trench (Supplementary Section S10). The $Sb_2S_3$ was then encapsulated by 40-nm-thick $Al_2O_3$ through thermal ALD (Oxford Plasmalab 80PLUS OpAL ALD) at 150 °C. To ensure good contact between the electric probe and metal pads while applying electrical pulses, the $Al_2O_3$ on the metal contacts was removed by defining a window using a sixth EBL with 600 nm PMMA, then etching in a chlorine-based inductively coupled plasma etcher (ICP-RIE, Oxford PlasmaLab 100 ICP-18).

*Optical simulation*

The refractive index data for $Sb_2S_3$ were measured by an ellipsometer (Woollam M-2000)[34]. The phase shifters and asymmetric directional couplers were designed (verified) by a commercial photonic simulation software package Lumerical MODE (FDTD).

*Heat transfer simulation*

The doped silicon PIN heater was simulated with a commercial Multiphysics simulation software COMSOL Multiphysics[35,36]. In the simulation, a heat transfer in a solid model is coupled with a semiconductor model to simulate the transient time performance.

*Optical transmission measurement setup*

The programmable units were measured with a 25°-angled vertical fiber-coupling setup. The stage temperature was controlled at 26°C by a thermoelectric controller (TEC, TE Technology TC-720). A tunable continuous-wave laser (Santec TSL-510) sent in the input light, the polarization of which was controlled by a manual fiber polarization controller (Thorlabs FPC526) to achieve a maximum fiber-to-chip coupling efficiency. A low-noise power meter (Keysight 81634B) measured the static optical transmission. The transmission spectra of all $Sb_2S_3$ devices were normalized to the spectra of the nearest reference waveguide. For the on-chip electrical switching, electrical pulses were applied to the on-chip metal contacts via a pair of electrical probes on two probe positioners (Cascade Microtech DPP105-M-AI-S). The crystallization and amorphization pulses were generated from a pulse function arbitrary generator (Keysight 81160A). The tunable laser, power meter, thermal controller, source meter, and pulse function arbitrary generator were controlled by a LabView program[35].

When electrically switching $Sb_2S_3$, we found that a single amorphization or crystallization voltage pulse switches the device to transmission levels with large state-to-state variations (Supplementary Section S11). This could be attributed to the switching of $Sb_2S_3$ into random intermediate structural phases since it has two (or more) crystalline phases[32]. This issue was tackled by using three identical pulses to switch the $Sb_2S_3$ phase completely.

We also note that sub-millisecond pulses failed to trigger the crystallization for the first time. We increased the voltage of a 500-µs pulse up to the material ablation voltage level without observing any crystallization. This suggests that the thermally induced $Sb_2S_3$ crystallization process is relatively slow, different from laser-induced crystallization[22], where a crystallization speed of tens of nano-second was demonstrated. We attribute this behavior to the difficulty in the initial nucleation. After the first crystallization process, consecutive crystallization process could happen at a scale of several hundreds of µs (still slow enough to avoid re-crystallization during the amorphization process, Supplementary Section S12). We hypothesize that crystallization becomes easier after the initial nuclei are formed by the first long thermal process. We note that we used 200-ms pulses in the reported experiments because this shorter pulse required after the first crystallization was found later. Therefore, the pulse condition in this work could be further optimized.

However, this limited crystallization speed is generally acceptable for many PIC applications, such as nonvolatile switching fabrics, optical programmable gate arrays, optical neural networks, where high speed is not essential and the zero-static power is more important[37,38]. Moreover, the slow crystallization speed could prevent unintentional recrystallization during long-pulse amorphization and permit a thicker layer of $Sb_2S_3$ to switch completely. We verified that a pulse with 10 µs duration was able to trigger a large degree of amorphization (Supplementary Section S13). This capability of actuating phase transition in thicker films is useful in photonics, since a much larger volume of PCMs are switched compared to electronic applications.

*Thermal parameter measurement*

The thermal conductivity of our films was measured with time-domain thermoreflectance (TDTR)[39,40]. TDTR uses ultrafast modulated laser heating through the absorption of a thin metallic transducer layer (70 nm Pt). An unmodulated probe laser then measures the surface temperature through a proportional change in the transducer reflectivity. Measurements were taken at a pump beam modulation frequency of 10 MHz to ensure the thermal penetration depth exceeded the thickness of the films. Knife edge measurements provided $1/e^2$ beam radii of 5.5 and 3.1 ± 0.05 µm for the pump and probe, respectively. The resulting thermoreflectance data were then fit to the solution of a 3D heat diffusion model for a multi-layer stack of materials (Pt-film-substrate) and the effective thermal resistance was determined as a function of film thickness. A linear regression of the foregoing data provided the thermal boundary resistance of the material stack, which was then used to determine the intrinsic thermal conductivities for the films. The properties of the Pt layer, the films, and the Si substrate were determined from independent measurements or adopted from the literature[41,42].

## Acknowledgments

The authors thank Asir Intisar Khan, Kathryn M. Neilson, Prof. Eric Pop at Stanford University, and Prof. Juejun Hu at MIT for the insightful discussions.

**Funding:** The research is funded by National Science Foundation (NSF-1640986, NSF-2003509), ONR-YIP Award, DARPA-YFA Award, NASA-STTR Award 80NSSC22PA980, and Intel. F.M. is supported by a Draper Scholars Program. Part of this work was conducted at the Washington Nanofabrication Facility/ Molecular Analysis Facility, a National Nanotechnology Coordinated Infrastructure (NNCI) site at the University of Washington, with partial support from the National Science Foundation via awards NNCI-1542101 and NNCI-2025489. Part of this work was performed at the Stanford Nano Shared Facilities (SNSF), supported by the National Science Foundation under award ECCS-1542152).

**Author contributions:** R.C. and A.M. conceived the project. R.C. simulated and fabricated the $Sb_2S_3$ silicon photonic devices, performed optical characterizations and data analysis. Z.F. and J.Z. helped with the device simulation, fabrication, characterization, and data analysis. F.M., S.G., K.K. and A.S. helped with the optical measurements. C.P. and K.E.G. measured thermal conductivity of $Sb_2S_3$ thin films. A.M. supervised and planned the project. R.C. wrote the manuscript with input from all the authors.

**Competing interests:** Authors declare that they have no competing interests.

**Data availability:** The data that support the findings of this study are available from the corresponding author upon reasonable request.

## Supplementary Information

# Supplementary information
# Non-volatile electrically programmable integrated photonics with a 5-bit operation

Rui Chen[1,*], Zhuoran Fang[1], Christopher Perez[3], Forrest Miller[1], Khushboo Kumari[1], Abhi Saxena[1], Jiajiu Zheng[1], Sarah J. Geiger[4], Kenneth E. Goodson[3], Arka Majumdar[1,2,*]

[1]Department of Electrical and Computer Engineering, University of Washington, Seattle, WA 98195, USA

[2]Department of Physics, University of Washington, Seattle, WA 98195, USA

[3]Department of Mechanical and Aerospace Engineering, Stanford University, Stanford, CA 94305, United States

[4]The Charles Stark Draper Laboratory, Cambridge, MA 02139, USA

*Email: charey@uw.edu and arka@uw.edu

**This supplementary information includes:**

S1. $Sb_2S_3$ thin film characterization

S2. Lumerical simulation for $Sb_2S_3$-SOI hybrid phase shifters and directional couplers

S3. Local crystal grains – micrograph image

S4. Extra measurement cyclability test results

S5. Multilevel performance based on $Sb_2S_3$ partial crystallization

S6. Heat transfer simulation

S7. Micrograph images for different intermediate levels

S8. The Table shows pulse conditions for 5-bit operation

S9. Performance retention after 77 days

S10. $Sb_2S_3$ stripe thickness after liftoff depends on the pattern width

S11. Single-pulse vs. multi-pulse switching

S12. Nucleation-speed limited initial crystallization and successive faster crystallization

S13. Long-pulse amorphization – evidence for large volume amorphization

## Section S1. $Sb_2S_3$ thin film characterization

We started with characterizing the thin film $Sb_2S_3$ using ellipsometry, X-Ray diffraction, and Raman spectroscopy (*Fig. S1*). A layer of 60-nm $Sb_2S_3$ was sputtered on a silicon wafer. The as-deposited a-$Sb_2S_3$ was characterized first and then switched to the crystalline phase by annealing at 325°C for 10 minutes under nitrogen flow.

The $Sb_2S_3$ was characterized by ellipsometry, and the data is fitted with Cody-Lorentz models with low mean square error (~5). The fitted complex refractive indices in *Fig. S1(a)* show a drastic change in the real part $n$ (~0.7 at 1310 nm), while almost zero $\kappa$ (0.0003 for amorphous and 0.03 for crystalline phases at 1310 nm) across the entire telecommunication O- and C-band. Therefore, switching the $Sb_2S_3$ produces a phase-only modulation. The microstructural phase transition after annealing and the stoichiometry were verified under X-ray diffraction (XRD) and Raman spectroscopy[1], as shown in *Fig. S1(b)* and *Fig. S1(c)*, by the characteristic lattice constants in the XRD and wavenumber shifts in the Raman spectrum. These characteristics show good agreement with existing literature[2].

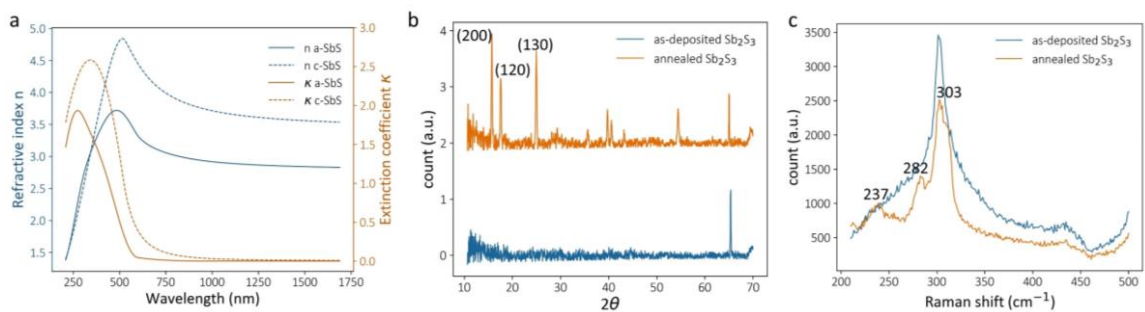

Fig. S1: $Sb_2S_3$ material characterization. (a) Broadband complex refractive index fitted from ellipsometry measurement. (b) X-ray diffraction and (c) Raman spectroscopy measurement for the as-deposited (blue) and annealed (orange) $Sb_2S_3$ samples. The peak at $2\theta = 65°$ for as-deposited $Sb_2S_3$ comes from the silicon substrate. The characteristic lattice constants and Raman shifts are marked.

# Section S2. Lumerical simulation for $Sb_2S_3$-SOI hybrid phase shifters and directional couplers

## S2.1. $Sb_2S_3$-based silicon phase shifters

*Fig. S2* shows the simulated $\pi$ phase shift length $L_\pi$ and insertion loss for the $Sb_2S_3$-based phase shifters. The refractive indices of $Sb_2S_3$ used here were obtained experimentally in the previous section. For a generic 500-nm-wide, 220-nm-high SOI waveguide with 20-nm-thick, 450-nm-wide $Sb_2S_3$ on the top, an effective index contrast of 0.018 is obtained at 1310 nm in *Figs. S2(a, b)*, indicating a $\pi$ phase shift length $L_\pi \approx 38\ \mu m$. *Figs. S2(c, d)* shows that a wider and thicker $Sb_2S_3$ film reduces $L_\pi$ and does not impact the insertion loss much (~0.23 dB/$\pi$). We adopted a generic waveguide width of 500 nm to avoid any extra taper structure, which renders $L_\pi \approx 38\ \mu m$. The calculated mode coupling between the bare silicon waveguide mode and the c-$Sb_2S_3$-Si hybrid waveguide mode is 99.7% (or -0.013 dB), which is negligible.

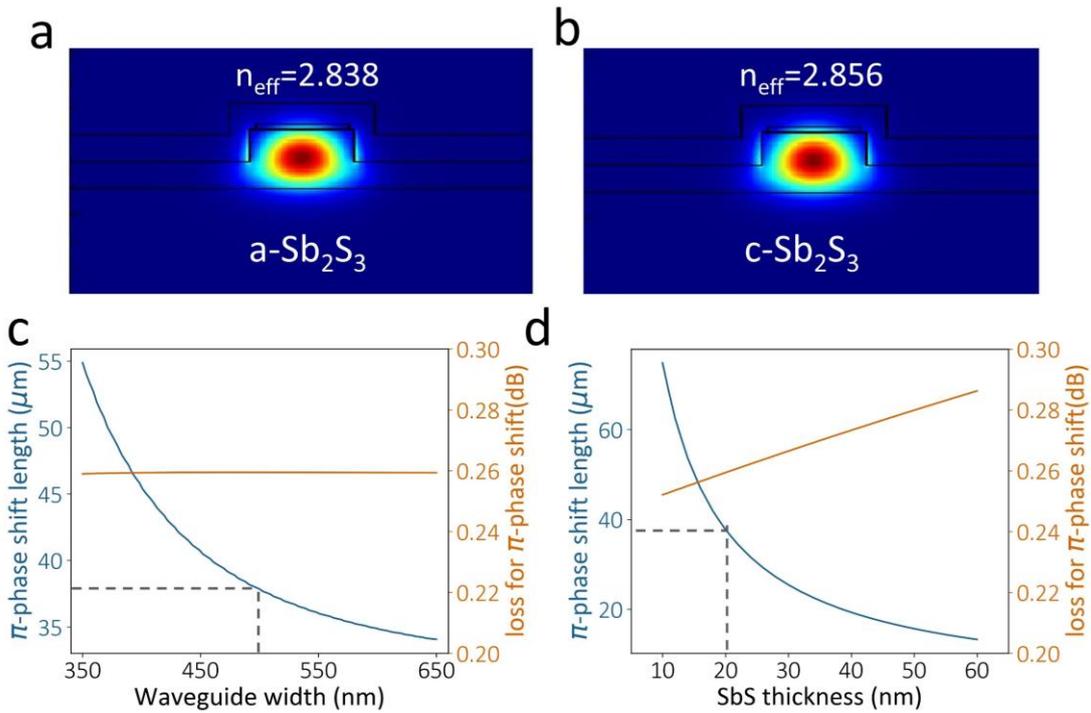

Fig. S2: Simulations for $Sb_2S_3$-on-SOI phase shifters. (a, b) Mode simulations for (a) amorphous and (b) crystalline $Sb_2S_3$. An effective index contrast of 0.018 is obtained at 1310 nm, indicating a $\pi$ phase shift length $L_\pi \approx 38\ \mu m$. (c, d) $L_\pi$ (blue) and excess loss for $\pi$ phase shift (orange) versus (c) the hybrid waveguide width and (d) $Sb_2S_3$ thickness. $L_\pi$ decreases as the increase of both waveguide width and $Sb_2S_3$ thickness, while the excess loss remains almost invariant. The dashed gray lines represent that we used a waveguide width of 500 nm and an $Sb_2S_3$ thickness of 20 nm in our experiments.

We note that $L_\pi$ could be further reduced by exploiting a wider or thicker $Sb_2S_3$. Although thick $Sb_2S_3$ gives a more compact device footprint, the complete phase transition is more difficult.

The vertical temperature gradient at thick $Sb_2S_3$ films could lead to $Sb_2S_3$ ablation at the bottom while not high enough temperature for amorphization at the top. Unintentional re-amorphization could also happen with a thick $Sb_2S_3$ layers. Moreover, recrystallization could also occur because of the crystallization kinetics and finite heat diffusion time[3]. To accommodate this tradeoff, we pick 20 nm as the experimental thickness, providing both a complete, repeatable $Sb_2S_3$ phase transition and a relatively compact $L_\pi$ of 38μm. Since this is the first experiment to switch $Sb_2S_3$ electrically, we chose a conservative thickness of 20 nm. Considering the good switching behavior observed in our experiment, a thicker $Sb_2S_3$ could be used in future experiments to provide even more compact devices. The largest thickness of $Sb_2S_3$ films that could be switched completely require more experimental exploration.

## S2.2. $Sb_2S_3$-based tunable asymmetric directional coupler

The asymmetric directional coupler is designed using the Lumerical Mode simulator and the coupled-mode theory. The idea has been depicted in the main text and the literature[4–6]. *Figs. S3 (a, b)* show two main optimization steps: (i). find the $Sb_2S_3$-loaded waveguide width with a fixed $Sb_2S_3$ thickness so that it is phase matched with a bare SOI waveguide; (ii). obtain a high transmission at the bar port when switching the $Sb_2S_3$ by optimizing the gap between two waveguides to allow the coupling length in the bar state to be an even multiple of the coupling length in the cross state[4]. Unlike the 1 × 2 coupler in ref[4], here we consider a 2 × 2 directional coupler, where light comes in from both input ports. To achieve an input port independent performance, the $Sb_2S_3$-SOI hybrid waveguide and the bare SOI waveguide are phase matched when $Sb_2S_3$ is in the crystalline phase. Intuitively, the light interacts with the slight loss of the $Sb_2S_3$ waveguide half of the $L_c$ regardless of the input port in this configuration.

After accomplishing the optimization in the Lumerical MODE simulator, we ran a Finite Difference Time Domain (FDTD) simulation to verify the design, and the results are in *Figs. S3 (c-f)*. The transmission spectra in *Figs. S3 (c, e)* show that a bar (cross) state is achieved by controlling the phase of $Sb_2S_3$ to amorphous (crystalline). At 1310 nm, the insertion loss is around 0.1 dB and 0.3 dB for the bar and cross state, respectively, and the extinction ratio is larger than 15 dB in both states. The higher insertion loss for the cross-state is mainly due to the slight material absorption of c-$Sb_2S_3$. The inset in *Fig. S3 (e)* shows that our device performance is symmetric and independent of the input port. *Figs. S3 (d, f)* shows the field propagation profile corresponding to *Figs. S3 (c, e)*. Again, we can verify an input-port independent performance.

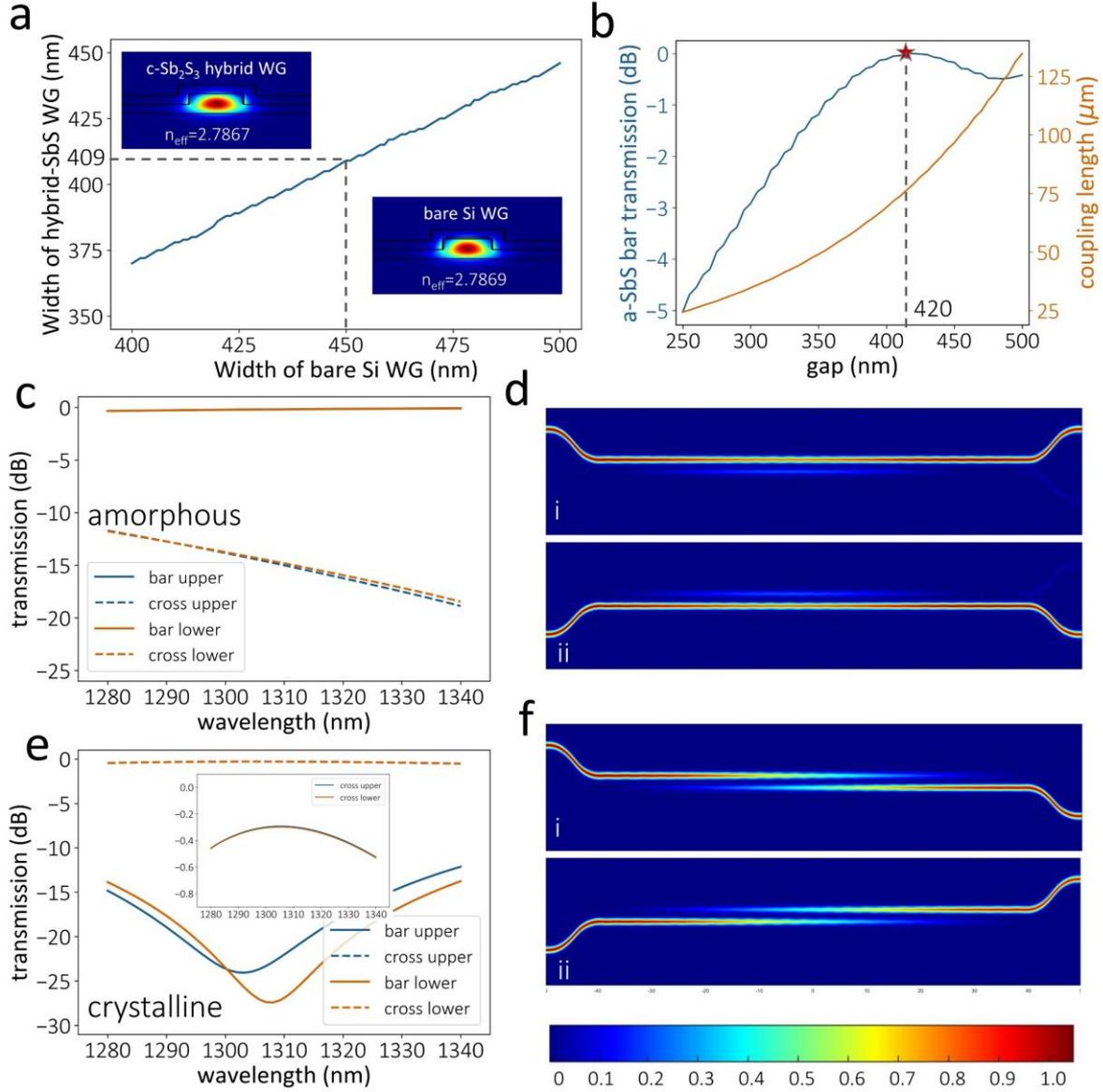

Fig. S3: Simulations for $Sb_2S_3$-on-SOI $2 \times 2$ tunable directional couplers with low insertion loss and large extinction ratios. (a) Phase-matching widths for c-$Sb_2S_3$-SOI and bare SOI waveguide. The dashed gray line shows that a 450-nm-wide bare SOI waveguide is phase matched with a 409-nm-wide c-$Sb_2S_3$-SOI waveguide. Their mode profiles are shown in the insets with a similar effective index, indicating good phase matching. (b) The gap (blue) is optimized to maximize bar port transmission when the $Sb_2S_3$ is switched into its amorphous phase. The coupling length $L_c$ for the cross state (orange) is also obtained based on the MODE simulation and coupled-mode theory. An optimal gap of 420 nm is picked, corresponding to $L_c \approx 79 \mu m$. (c - f) FDTD simulation results. (c, e) Transmission spectrum and (d, f) field profile for (c, d) amorphous- and (e, f) crystalline-$Sb_2S_3$. (i) and (ii) show the field profile when inputting light from the upper and lower port, respectively. The inset in (e) shows an input-port independent performance for the cross-state.

On the contrary, if the waveguides are phase matched for a-$Sb_2S_3$, the loss would be much higher when light inputs from the $Sb_2S_3$ waveguide. *Fig. S4* shows the simulated spectrum of an optimized design, where 20-nm a-$Sb_2S_3$ is phase-matched with a bare SOI waveguide. As shown in the zoomed-in bar transmission plot *Fig. S4 (b)*, a difference of 0.3 dB between two input ports is observed when $Sb_2S_3$ is in its crystalline phase. This seemingly slight asymmetry could lead to unintentionally unbalanced optical paths in a large-scale PIC system.

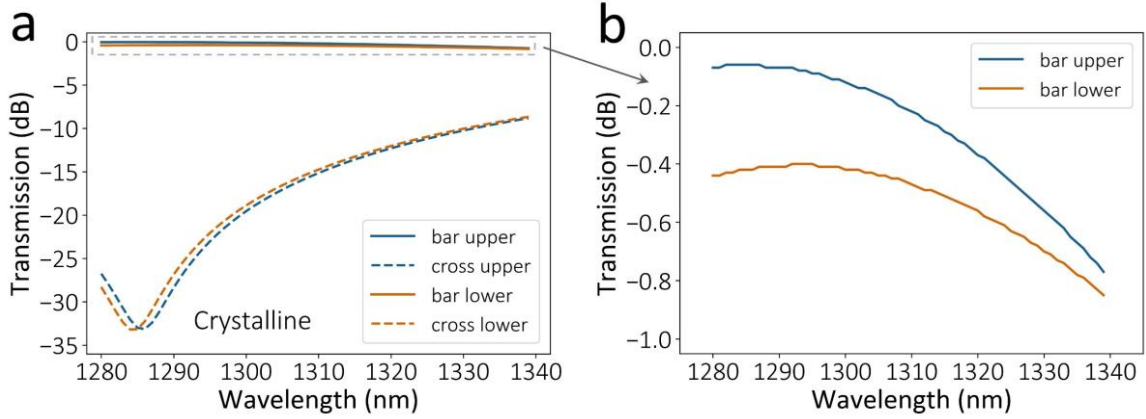

Fig. S4: Phase matching a-$Sb_2S_3$-SOI waveguide with bare SOI waveguide results in an asymmetric (input port dependent) bar state when switching the $Sb_2S_3$ to its crystalline phase. (a) Transmission spectrum for c-$Sb_2S_3$. (b) Zoomed in transmission spectrum at the bar port shows a transmission difference of 0.3dB between two input ports.

We adopted the relatively thin $Sb_2S_3$ with 20 nm thickness to ensure a complete phase transition. However, such a thin PCM layer provides limited tunability, resulting in a large device footprint. Fortunately, we note the slow crystallization speed of $Sb_2S_3$ could potentially enable a much thicker PCM layer to crystallize (see Section S2.1), which will shrink the device size. Using the same design methodology, we numerically designed asymmetric directional couplers with $Sb_2S_3$ thickness ranging from 30 nm to 50 nm in *Table. S1*. Remarkably, the designed directional coupler with 50-nm-thick $Sb_2S_3$ shows a significantly reduced coupling length (~34µm).

Table. S1: Performances of $Sb_2S_3$ directional couplers using different $Sb_2S_3$ thicknesses

| $t_{SbS}$ (nm) | $W_b$ (nm) | $W_h$ (nm) | gap (nm) | $L_c$ (µm) |
|---|---|---|---|---|
| 20 | 450 | 409 | 420 | 79 |
| 30 | 450 | 395 | 370 | 55 |
| 40 | 450 | 383 | 330 | 41 |
| 50 | 450 | 373 | 305 | 34 |

Note: $t_{SbS}$, $W_b$, and $W_h$ represent the $Sb_2S_3$ thickness, and widths of bare SOI and hybrid $Sb_2S_3$-SOI waveguide, respectively.

## S2.3. Estimate loss of asymmetric directional coupler using experimentally extracted $Sb_2S_3$ loss

We extract the actual loss of c-$Sb_2S_3$, including the crystal grain scattering loss, by match the loss measured in the high-Q ring resonator (0.76 dB/$\mu$m) with MODE simulation. We obtained an extinction coefficient $\kappa' \approx 0.015$, compared to ellipsometry measurement result of $\kappa = 0.005$ at a wavelength of 1310 nm. We then estimate the actual loss of the asymmetric directional coupler using the new complex refractive index. *Fig. S5* shows the simulation results, where the insertion loss is around 0.85 dB and the extinction ratio is 25 dB. We note the amorphous $Sb_2S_3$ loss is very small in the experiment, so the initial simulation result still holds (0.1 dB insertion loss and 15 dB extinction ratio).

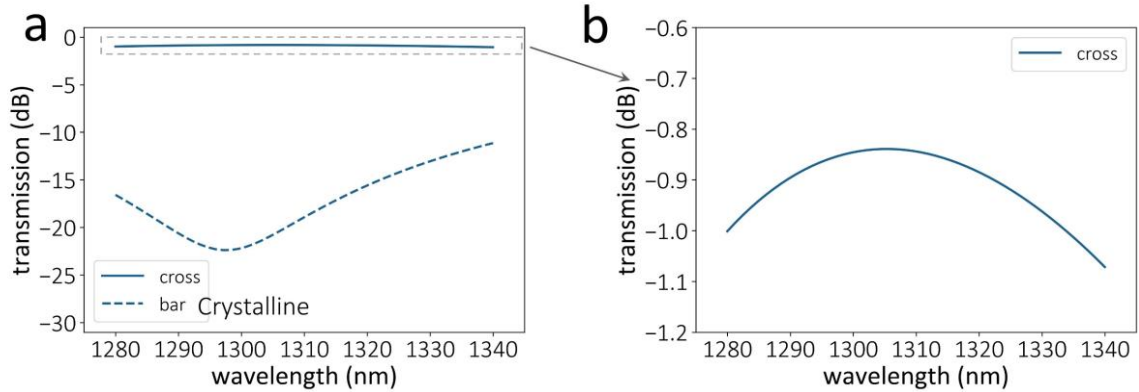

Fig. S5: Transmission spectrum for asymmetric directional coupler using the ring resonator measured c-$Sb_2S_3$ loss. The insertion loss increases from 0.3 dB to 0.85 dB and the extinction ratio does not change significantly.

## S2.4. A multimode interferometer (MMI)

A 120-nm partially etched MMI was designed using Lumerical Eigen-Mode Expansion (EME) Method. The design was further tweaked and verified with 3D FDTD simulations. The optimized multimode device parameters are annotated in *orange in Fig. S6 (a)*. *Figs. S6 (a, b)* show the simulated electric field propagation profile and the transmission spectra at both output ports. The optimized MMI exhibits an insertion loss < 0.1 dB at 1310 nm and a 1-dB bandwidth > 60 nm, as shown in *Fig. S6 (b)*. The measured transmission spectra are presented in *Fig. S6 (c)*. The slight spectral fluctuation (~0.1 dB) is attributed to the interference between back-reflected light at the interface between the single mode tapers and the multimode region.

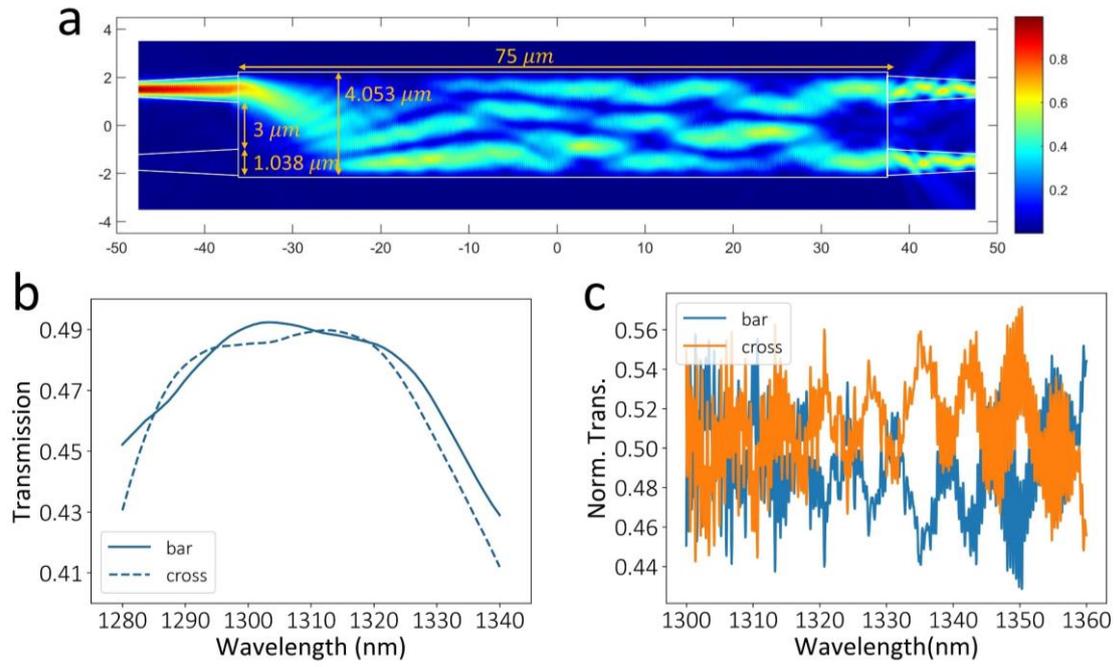

Fig. S6: Low-loss 50:50 multimode interferometer (MMI) at 1310 nm. (a) Field profile and (b) transmission spectrum using FDTD simulation. The white dotted line outlines the device shape. (c) Measured transmission spectrum of the fabricated device.

## Section S3. Local crystal grains – micrograph image

In *Fig. S7*, we took a micrograph image of 60-nm-thick $Sb_2S_3$ on a blanket silicon chip, where the $Sb_2S_3$ films were annealed under 325°C and in the crystalline phase. The micrograph image shows a non-uniform surface with a lot of local crystal grains and even flower-shaped patterns (circled in red), which could lead to unintended scattering loss for in-plane light propagation. We hypothesize that the crystal grains form because crystalline $Sb_2S_3$ has a density reduction of around 24% compared to its amorphous form[7]. Hence, a significant strain is induced during the phase transition, and results in a non-uniform surface. We also speculate that the flower-like patterns occurs near the nuclei.

It is unclear what the c-$Sb_2S_3$ surface looks like on our devices because those 20-nm 450-nm-wide $Sb_2S_3$ films are too small to image optically. The 40-nm-thick encapsulation alumina further prevents the accurate usage of SEM. Further material study is required to quantitatively understand the crystalline patterns formed.

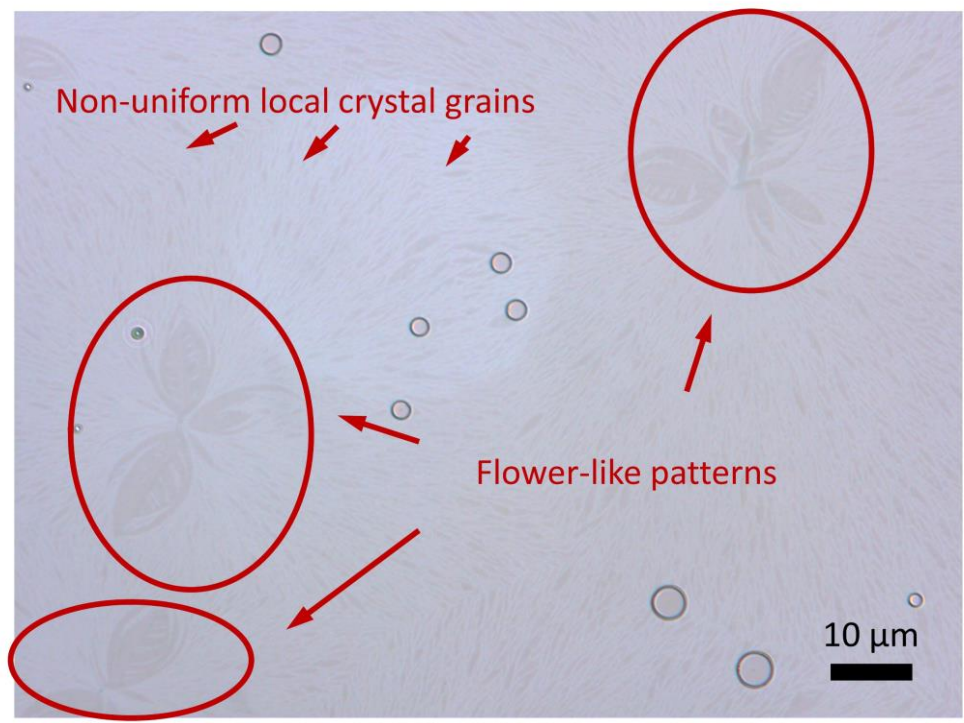

Fig. S7: Non-uniform surface forms after the annealing due to the density difference between amorphous and crystalline $Sb_2S_3$. Some flower-shaped patterns are observed. We speculate the nonuniform surface is due to the large density difference between a- and c-$Sb_2S_3$, and the flower-shaped patterns occur near the nuclei.

## Section S4. Extra cyclability test results

**MZI cyclability result – 100 cycles**

The $Sb_2S_3$ phase shifter in an MZI was switched reversibly by alternatively applying SET and RESET pulses. We note this cyclability test was performed with a single pulse scheme, so a significant variation was observed compared to *Fig. 4* in the main text, where three pulses were used for each amorphization or crystallization. Yet, we found 100 consecutive switching events (*Fig. S8*) where the device was switched with a high extinction ratio of ~15 dB.

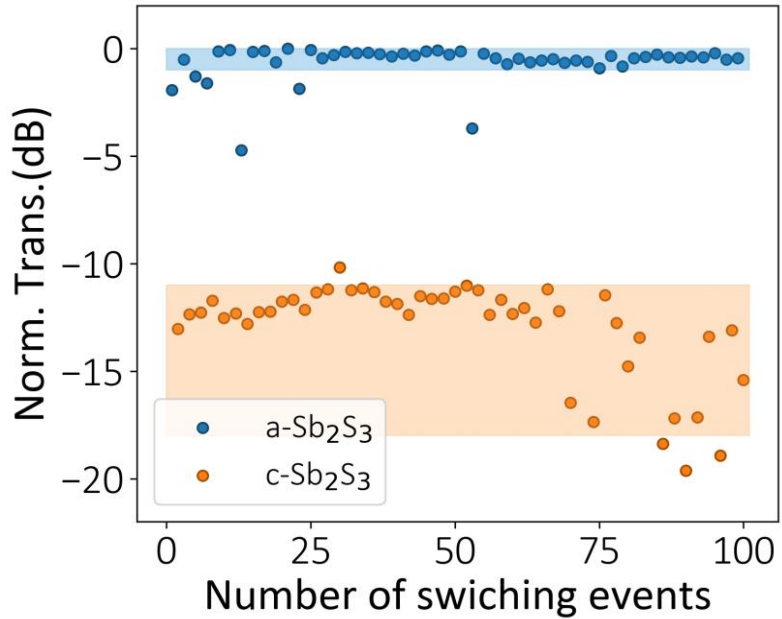

Fig. S8: The $Sb_2S_3$-based phase shifter in an MZI is experimentally switched between amorphous (blue) and crystalline (orange) phases for 100 switching events. Each switching event was triggered with a single pulse.

**Asymmetric directional coupler transmission before and after 2,300 switching events**

We switched the asymmetric directional coupler for 2,300 switching events. However, some drift in the optical measurement setup occurred, so only 1,600 events were reported in the main text. In this section, we compare the SEM image and the transmission spectrum before and after 2,300 switching events, and give an explanation of the performance degradation or the transmission drifting.

*Figs. S9 (a, b)* show the scanning electron-beam micrograph images of $Sb_2S_3$ directional couplers before and after the cyclability test. We note that the black dots on the lower waveguide (*Fig. S9 (a)*) is likely the crystalline $Sb_2S_3$ nuclei formed during the rapid thermal annealing. The rough $Sb_2S_3$ edge in *Fig. S9 (b)* indicates that switching the relatively large volume of the $Sb_2S_3$ stripe many times results in a distorted $Sb_2S_3$. This is because the thin film always tends to minimize the surface energy during melt-quench process, hence leads to sinusoidal edge. This phenomena is known as Rayleigh instability of fluid motions[8]. The result of such surface-energy-induced instability could be discrete circular patterns, having been used for advanced nano-fabrication[9]. Therefore, to further improve the endurance of our devices, one possible approach is to pattern the long $Sb_2S_3$ stripes to subwavelength gratings to ensure a lower initial surface energy and a low excess loss. We note that this subwavelength grating idea has been experimentally demonstrated in other works using GST[10–12], hence could be a future direction to improving the presented work.

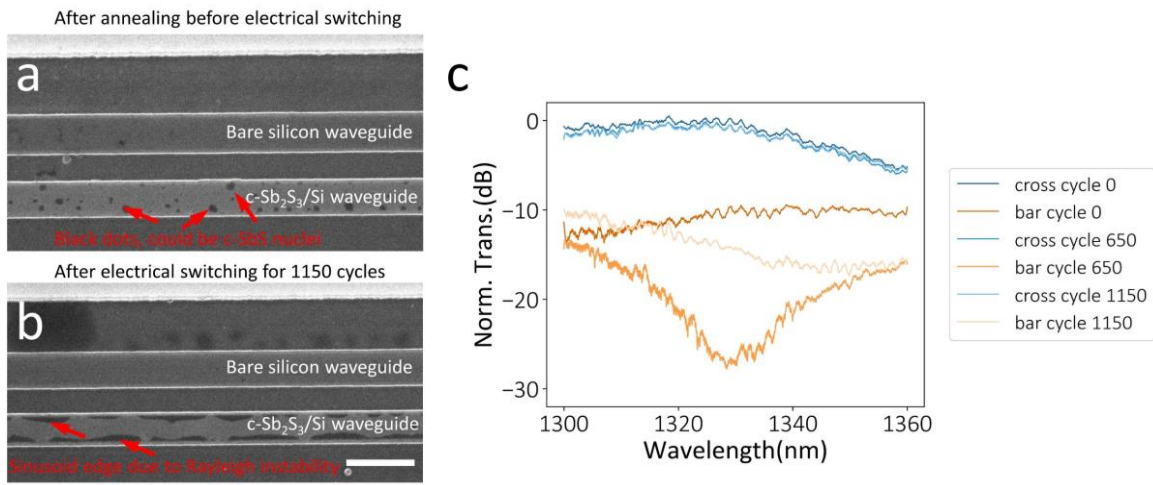

Fig. S9: Only slight device degradation occurred after 1,150 cycles (2,300 switching events), and the SEM images suggest it is due to thermal reflowing. (a, b) SEM images of a device after (a) a single rapid thermal annealing and (b) 1,150 cycles. Scale bar: 1 µm. (c) Transmission spectra before, in the middle, and at the end of the cyclability test. No significant performance degradation in the cross port (blue) was observed after switching the asymmetric directional coupler for more than 1,150 cycles. The variation in the bar port could be attributed to the change in $Sb_2S_3$ distribution due to the thermal reflowing or Rayleigh instability issue.

The transmission spectra before and after the cyclability test are shown in *Fig. S9 (c)*. A minor drift in the spectra could be attributed to the thermal reflowing, which changes the $Sb_2S_3$ shape, hence breaking the phase matching condition. Nevertheless, the device still shows a low insertion loss (< 1dB degradation) and high extinction ratio (~ 15 dB).

## Section S5. Multilevel performance based on Sb$_2$S$_3$ partial crystallization

By sending in relatively low voltage crystallization pulses (voltage 3.1 V, duration 200 ms, leading-edge 10 ms, falling edge 100 ms), we obtained partial crystallization performance in *Fig. S10*. Such behavior could be explained by the growth-dominant nature of Sb$_2$S$_3$[4]. The growth rate could be controlled by the temperature (pulse voltages); the total crystallization volume could be controlled with the duration of crystallization pulses. Further engineering the pulse parameters could potentially achieve more operation levels.

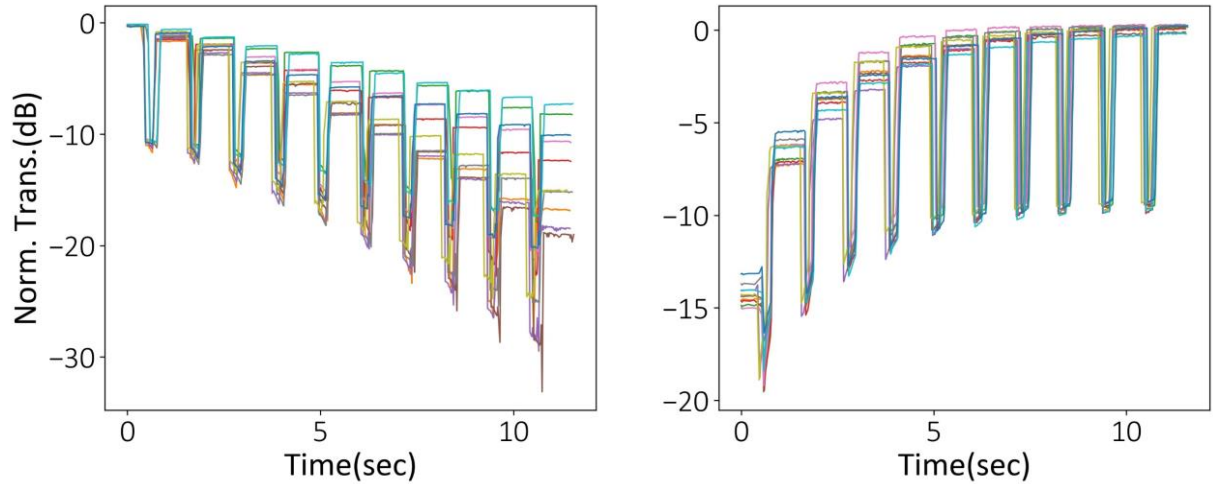

Fig. S10: Stepwise partial crystallization using identical, low amplitude, and long duration pulses with a one-second interval. Continuous time trace measurement at (a) bar and (b) cross port of an asymmetric directional coupler. The experiment was repeated for 10 times at each output port, represented by different colors, and the multilevel characteristic occurred in all experiments.

## Section S6. Heat transfer simulation results

A heat transfer simulation in COMSOL Multiphysics is presented in this section[6,13–16]. The large thermal gradient between $Sb_2S_3$ and its surroundings leads to a fast cooling rate. Such a fast cooling rate ($> 10^9\ K/s$ [17]) is necessary for complete amorphization to take place. Simulation shows that the temperature returns to near room temperature within a few $\mu s$ (*Fig. S11*). The $\mu s$-level thermal relaxation time rules out any residual heat when we applied the one-second interval electrical pulses for stepwise multilevel amorphization.

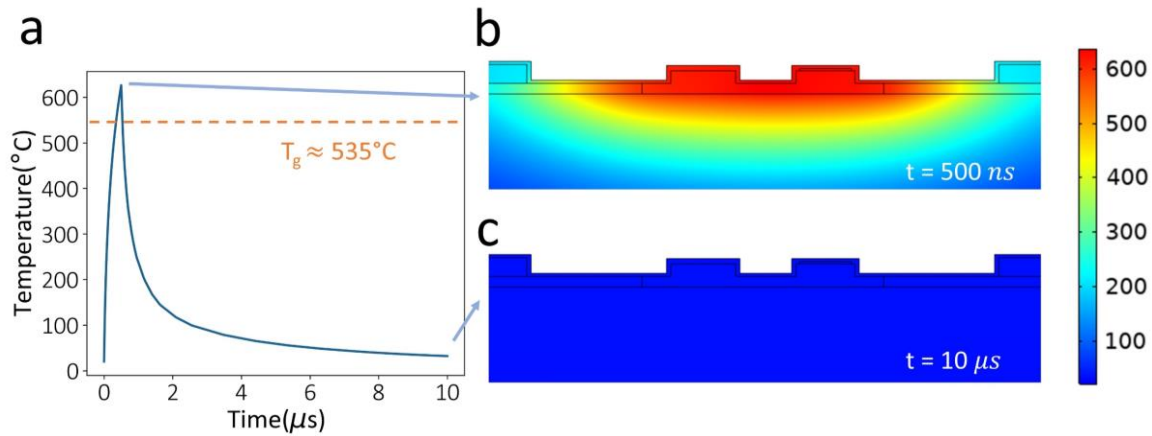

Fig. S11: The thermal relaxation time for a 500 ns amorphization pulse is around several $\mu s$. (a) Time-dependent heat transfer simulation results at the $Sb_2S_3$ thin film. (b, c) The thermal profiles at (b) t = 500 ns and (c) t = $10\mu s$.

## Section S7. Micrograph images for different intermediate levels

We inspected devices at different intermediate levels under the microscope in *Fig. S12*. One can see a gradual increase of amorphous $Sb_2S_3$ as the degree of amorphization increases. The amorphous $Sb_2S_3$ is not located precisely in the center of the long strip, where the temperature is the highest. We attribute this to the possible non-uniform doping profile and recrystallization of $Sb_2S_3$ near the center.

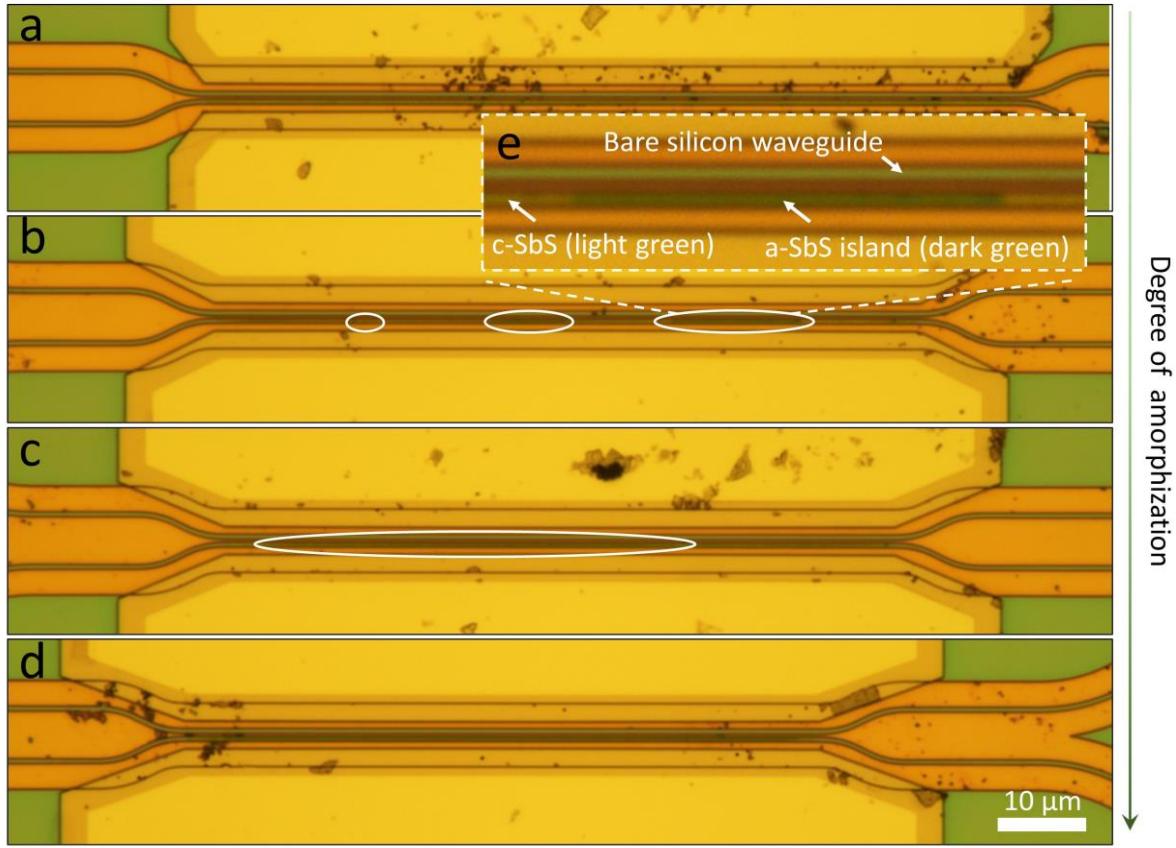

Fig. S12: Micrograph images for directional couplers in different intermediate levels. The levels are set using identical pulses with one-second intervals. The degree of amorphization increases from (a) to (d). The dark (light) green regions on the lower waveguide are a(c)-$Sb_2S_3$. Islands of a-$Sb_2S_3$ (circled in red) are visible in (b) and (c) and continue to grow. (e) A zoom-in picture of the right-most circled region in (b). A clear boundary between a- and c-$Sb_2S_3$ is observed in the $Sb_2S_3$/Si hybrid waveguide (the lower waveguide).

# Section S8. Table shows pulse conditions for 5-bit operation

We engineered the pulse condition dynamically to achieve the high 0.5-dB precision. We started with low voltage (9.65 V) that could barely trigger the amorphization and gradually increase the amplitude. In the meantime, we monitored the transmission after sending each pulse. If the transmission is not changed, we increase the voltage slightly (by 0.05 V). If the transmission changes, but smaller than (0.5 ± 0.1) V (this value is our target), we continue to send more pulses until the targetted transmission is obtained. If the transmission is changed by larger than (0.5 + 0.1) V, we partially crystallize the device and repeat the previous steps. We note that it is necessary to monitor the transmission and dynamically change the pulse conditions to mitigate the stochastic nature of PCMs[18]. The following Table shows the pulse conditions in the 5-bit operation experiments.

Table. S2: **Pulse conditions for 5-bit operation in five experiments (Unit: Volts).** Pulses all have a duration of 550 ns, but the amplitude is increasing gradually. Either single or multiple pulses were applied for each level.

| Level | Exp 1 | Exp 2 | Exp 3 | Exp 4 | Exp 5 |
|---|---|---|---|---|---|
| 1 | 9.7 | 9.65 | 9.75 | 9.6 | 9.6 |
| 2 | 9.7, 9.8 | 9.8 | $9.75 \times 3$ | 9.6 | 9.62, 9.63, 9.64 |
| 3 | 9.8 | 9.8 | 9.8 | $9.6 \times 2$ | 9.65 |
| 4 | 9.75 | 9.75 | 9.8 | 9.65 | 9.65, 9.62 |
| 5 | 9.75 | 9.75 | 9.8 | 9.65 | 9.62 |
| 6 | $9.75 \times 2$ | 9.75 | 9.8 | 9.65 | 9.65 |
| 7 | 9.7, 9.75 | $9.75 \times 2$ | 9.8 | 9.65, 9.6 | 9.65 |
| 8 | 9.75, 9.72 | 9.75 | 9.8 | 9.7 | 9.65 |
| 9 | $9.72 \times 4$ | 9.75 | 9.75 | 9.7 | 9.65, 9.63 |
| 10 | $9.72 \times 3$ | 9.75 | $9.75 \times 2$ | 9.67 | $9.67 \times 2$ |
| 11 | 9.75, $9.73 \times 2$ | 9.75 | 9.75 | 9.67 | 9.69 |
| 12 | $9.73 \times 3$ | 9.75 | $9.75 \times 2$ | 9.67 | 9.7, 9.67 |
| 13 | $9.73 \times 3$ | $9.75 \times 2$ | 9.8 | 9.67 | 9.72 |
| 14 | $9.73 \times 3$ | 9.75 | 9.75 | 9.68, 9.66 | $9.72 \times 2$ |
| 15 | $9.73 \times 3$ | $9.75 \times 2$ | 9.8 | 9.68, 9.67 | 9.72 |
| 16 | $9.74 \times 3$ | $9.75 \times 2$ | 9.75 | 9.69 | 9.72 |
| 17 | 9.76, 9.8 | $9.75 \times 2$ | 9.8 | 9.69, 9.7 | 9.72 |
| 18 | $9.8 \times 2$ | 9.8 | 9.8, 9.75 | 9.7, $9.67 \times 2$ | 9.72, 9.7 |
| 19 | $9.8 \times 2$ | 9.8 | 9.75 | 9.7 | 9.7 |
| 20 | $9.8 \times 3$ | 9.8 | 9.8 | 9.7 | 9.72 |
| 21 | $9.8 \times 2$ | 9.8 | 9.8, 9.75 | 9.7 | 9.73 |
| 22 | $9.8 \times 2$ | $9.8 \times 2$ | 9.8 | $9.7 \times 3$ | $9.75 \times 2$ |

| 23 | 9.8 × 2 | 9.8 | 9.8 | 9.73 | 9.75 × 2 |
| 24 | 9.85 | 9.8 × 2 | 9.8 | 9.73 | 9.78 |
| 25 | 9.85 | 9.85 | 9.85 | 9.75 | 9.8 × 2 |
| 26 | 9.85 | 9.85 | 9.8 | 9.75 | 9.8 |
| 27 | 9.85 | 9.85 | 9.85 | 9.75 | 9.82 |
| 28 | 9.85 | 9.85 | 9.85 | 9.75 × 2 | 9.85 |
| 29 | 9.85 | 9.85 | 9.85 | 9.77 | 9.8 |
| 30 | 9.85 | 9.85 | 9.8 | 9.77 × 3 | 9.85 |
| 31 | 9.85 | 9.85 | 9.85 | 9.77 × 3 | 9.85 |
| 32 | 9.85 | 9.85 | 9.85 × 2 | 9.77 × 3 | 9.85 |

## Section S9. Performance retention after 77 days

Our device is non-volatile under the ambient environment. We measured a device after leaving it in the ambient (open in the air) for 77 days and compared its performance in *Fig. S13*. The device retains its performance excellently, with slight variation, which the small difference in the grating coupler coupling condition (coupling angle) might cause.

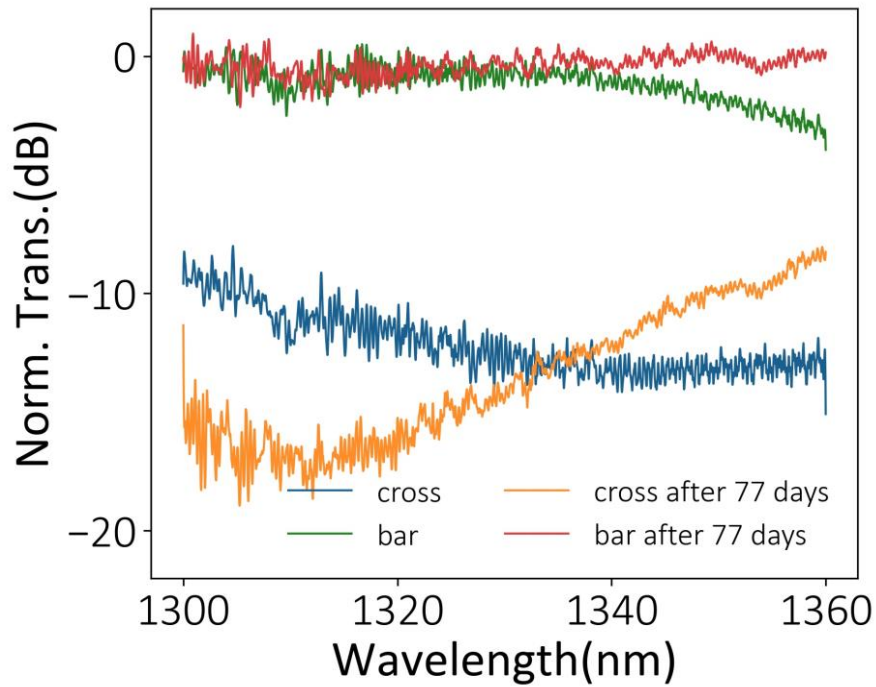

Fig. S13: An asymmetric directional coupler's characteristics was retained over 77 days.

# Section S10. Sb$_2$S$_3$ stripe thickness after liftoff depends on the pattern width

We find that a narrow resist trench for Sb$_2$S$_3$ deposition and liftoff leads to a reduced Sb$_2$S$_3$ thickness than blanket Sb$_2$S$_3$ deposition. This could be attributed to that part of Sb$_2$S$_3$ was blocked by the resist due to a deep trench. We fabricated a chip with different patch width (300, 400, 500 nm) and measure the thickness after Sb$_2$S$_3$ liftoff. The measured results (*Table. S3*) shows a reduction factor of around 0.5 compare to blanket deposition. We also measured the widths of the Sb$_2$S$_3$ widths which agree reasonably well with the Ebeam defined widths. Since this is not the main interest of this measurement, we only measured part of the Sb$_2$S$_3$ stripes, and the ones not measured are denoted as '-'.

Table. 3: Sb$_2$S$_3$ thickness reduces after liftoff measured by AFM (Unit: nm)

| Deposited width | Width after Liftoff | Deposited thickness | Thickness after liftoff |
|---|---|---|---|
| **300** | 373 | 30 | 13.8 |
| **300** | 354 | 30 | 14 |
| **400** | 471 | 30 | 16.9 |
| **400** | 471 | 30 | 17 |
| **300** | - | 40 | 16.4 |
| **400** | 413 | 40 | 19.5 |
| **300** | - | 50 | 21 |
| **400** | - | 50 | 26 |
| **400** | 471 | 50 | 25.9 |

## Section S11. Single-pulse vs. multi-pulse switching

In our experiment, single pulses could not trigger a reliable, complete phase transition. We sent in a single pulse for SET/RESET. The final state shows a significant variation (*Fig. S14 (a)*). This could be attributed to the multiple crystalline phases and incomplete thermal process due to the relatively thick $Sb_2S_3$. We tackled this issue by sending two or three pulses to trigger a complete phase transition. The resulting transmission spectra of another ring resonator are shown in *Fig. S14 (b)*. This multi-pulse switching scheme achieved a much more consistent performance across ten cycles. We emphasize that this behavior is distinctly different compared to GST or $Sb_2Se_3$, where we were able to actuate the phase transition under a single shot, in a very similar PIC. While we provided a hypothesis for such behavior, more material studies are warranted to explain the behavior.

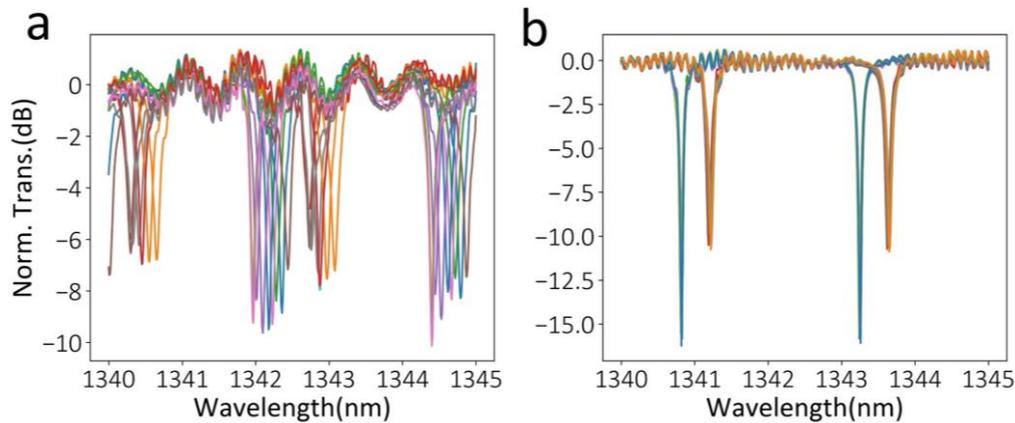

Fig. S14: Single SET (RESET) pulses could not reliably trigger complete crystallization or amorphization. (a, b) Transmission spectra of a ring resonator switched with (a) single SET(RESET) pulses and (b) three identical SET(RESET) pulses with one-second intervals. Different colors represent different experiments.

## Section S12. Nucleation-speed limited initial crystallization and successive faster crystallization

Experimentally, we observed a very slow $Sb_2S_3$ crystallization for the first time, where even 500 μs pulses could not trigger the crystallization. The devices were then crystallized with DC voltage or pulses with very long durations such as 200 ms. This could be explained by the small enthalpy of fusion of $Sb_2S_3$ (~40.64 kJ/mol[19], compared to fusion enthalpy[20] of GST 625 J/cm$^3$ or $\frac{625}{\rho}M = \frac{625\ J/cm^3}{5.87\ g/cm^3} \cdot 1026.8\ g/mol = 109.3\ kJ/mol$, where $\rho$ is the density[21] and $M$ is the molar mass of GST), which necessitates a large critical nuclei size to overcome the crystallization energy barrier.

After the first crystallization, however, the successive crystal growth process happens without any requirement for overcoming the energy barrier. We were able to see partial crystallization using much shorter 100-μs pulses, as shown in Fig. S15. We note this slow crystallization guarantees larger volumes of $Sb_2S_3$ amorphized again without recrystallization and is thus could be beneficial for photonic applications. In the next section, we experimentally show that indeed the amorphization could be triggered with very long pulses.

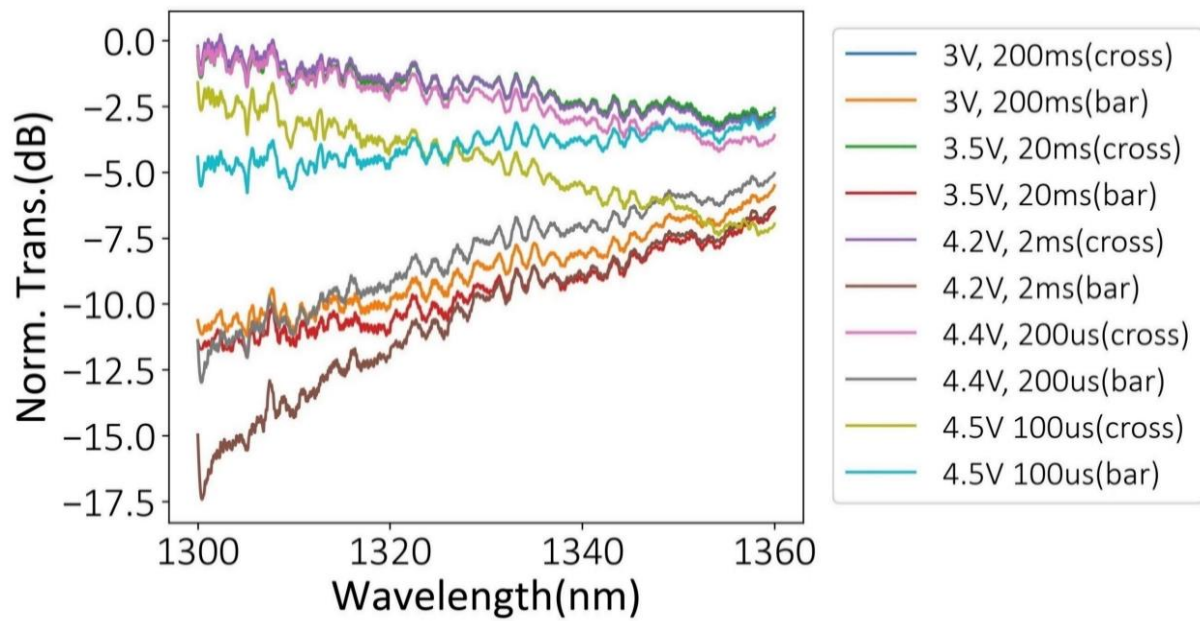

Fig. S15: The slow crystallization nature of $Sb_2S_3$ limits the pulse duration longer than 100 $\mu s$. The 100-$\mu s$-long pulses were not able to trigger complete crystallization even after the first time. We note again that during the first few cycles of crystallization, only pulses with even much longer duration, such as 200 ms, could trigger the crystallization. This indicates the difficulty in forming initial $Sb_2S_3$ nuclei and the substantially faster rate of crystal growth compared with nucleation.

# Section S13. Long-pulse amorphization – evidence for large volume amorphization

One appealing aspect of using "slow" PCMs, such as $Sb_2S_3$, is that the slow crystallization process could avoid any unintentional recrystallization during the melt-quench process, hence could ensure the amorphization for a large volume of PCM. This capability of completely switching large volumes of PCMs is crucial in photonics, where the volume (typically orders of µm$^3$) are much larger than in electronic applications (typically orders of (10 nm)$^3$) and takes precedence over the phase transition speed[22].

Here, we verify $Sb_2S_3$'s resistance to unintentional recrystallization by comparing the longest amorphization time of an asymmetric directional coupler to other reported PCM devices. *Fig. S16* shows that the $Sb_2S_3$ on the directional coupler could be amorphized under different pulse durations, ranging from 500 ns to 10 µs. A large degree of amorphization was observed even when we increased the pulse durations to 10 µs. This 10-µs pulse duration is much longer than in the reported GST[6,14,23] (< 200 ns) or SbSe[24] (< 1 µs, but the thickness is 30 nm) switching experiments. This successful amorphization with long pulses supports that $Sb_2S_3$ is inherently more suitable for large volume amorphization because recrystallization is surpressed.

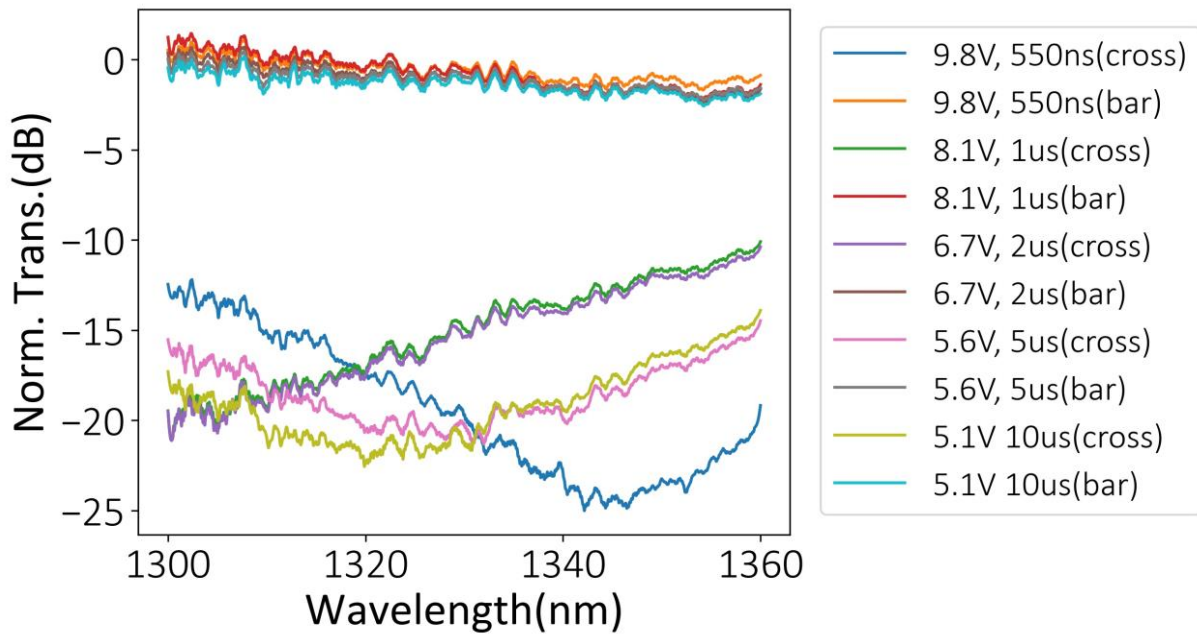

Fig. S16: Complete amorphization was triggered with relatively long, 10-$\mu s$ pulses because of the slow crystallization nature of $Sb_2S_3$. High-level port (bar port) transmission is consistent among all the pulse conditions. The low-level port (cross port) transmission variation could be attributed to a tiny portion of the $Sb_2S_3$ not being switched completely, which could be potentially overcome by optimizing the heater design.